%% file: main.tex
\documentclass[reprint,aps,physrev,superscriptaddress]{revtex4-2}

\include{preamble}

\begin{document}

\title{Dynamics of a dispersively coupled transmon qubit in the presence of a noise source embedded in the control line}
\date{\today}

\include{authors}
\input{abstract}
\maketitle

\input{introduction}
\input{model_derivation}

\input{solving_models}
\input{conclusions}
\input{acknowledgements}

\appendix
\input{appendix1}
\input{appendix2}
\input{appendix3}
\input{appendix4}
\input{appendix5}
\input{appendix6}

\bibliography{bibliography}

\end{document}

%% file: preamble.tex
\usepackage{amsmath}
\usepackage{amssymb}
\usepackage[utf8]{inputenc}
\usepackage[T1]{fontenc}
\usepackage{graphicx}
\usepackage{dcolumn}
\usepackage{bm}
\usepackage{mleftright}
\usepackage{bbold}
\usepackage{braket}
\usepackage{relsize}
\usepackage{hyperref}
\hypersetup{
  colorlinks   = true, 
  urlcolor     = blue, 
  linkcolor    = blue, 
  citecolor   = blue 
}
\usepackage{booktabs}

\usepackage{mathptmx}

\usepackage{tikz}
\usepackage[european]{circuitikz}
\usetikzlibrary{matrix}

\newcommand\restr[2]{{
  \left.\kern-\nulldelimiterspace 
  #1 
  \vphantom{\big|} 
  \right|_{#2} 
  }}

\def\e{\ensuremath{\mathrm{e}}}
\def\i{\ensuremath{\mathrm{i}}}
\def\d{\ensuremath{\mathrm{d}}}

\definecolor{myblue}{RGB}{0,80,158}
\definecolor{myred}{RGB}{125,0,45}

\DeclareMathOperator{\Tr}{Tr}
\def\liouv{\ensuremath\mathcal{L}}
\newcommand{\kket}[1]{|#1\rangle\rangle}
\newcommand{\kketbig}[1]{\big|#1\big\rangle\big\rangle}
\newcommand{\bbra}[1]{\langle\langle#1|}
\newcommand{\bbraket}[2]{\langle\langle #1 | #2\rangle\rangle}
\newcommand{\bbraketbig}[2]{\big\langle\big\langle #1 \big| #2\big\rangle\big\rangle}

%% file: authors.tex
\author{Antti Vaaranta}
\affiliation{Bluefors Oy, Arinatie 10, 00370 Helsinki, Finland
}
\email{antti.vaaranta@helsinki.fi}
\affiliation{QTF Centre of Excellence, Department of Physics, University of Helsinki, P.O. Box 43, FI-00014 Helsinki, Finland}

\author{Marco Cattaneo}
\affiliation{QTF Centre of Excellence, Department of Physics, University of Helsinki, P.O. Box 43, FI-00014 Helsinki, Finland}
\affiliation{Instituto de F{\'i}sica Interdisciplinar y Sistemas Complejos (IFISC, UIB-CSIC), Campus Universitat de les Illes Balears E-07122, Palma de Mallorca, Spain}

\author{Russell E.~Lake}
\affiliation{Bluefors Oy, Arinatie 10, 00370, Finland
}

%% file: abstract.tex
\begin{abstract}

We describe transmon qubit dynamics in the presence of noise introduced by an impedance-matched resistor ($50\,\mathrm{\Omega}$) that is embedded in the qubit control line. To obtain the time evolution, we rigorously derive the circuit Hamiltonian of the qubit, readout resonator and resistor by describing the latter as an infinite collection of bosonic modes through the Caldeira-Leggett model. Starting from this Jaynes-Cummings Hamiltonian with inductive coupling to the remote bath comprised of the resistor, we consistently obtain the Lindblad master equation for the qubit and resonator in the dispersive regime. We exploit the underlying symmetries of the master equation to transform the Liouvillian superoperator into a block diagonal matrix. The block diagonalization method reveals that the rate of exponential decoherence of the qubit is well-captured by the slowest decaying eigenmode of a single block of the Liouvillian superoperator, which can be easily computed. The model captures the often used dispersive strong limit approximation of the qubit decoherence rate being linearly proportional to the number of thermal photons in the readout resonator but predicts remarkably better decoherence rates when the dissipation rate of the resonator is increased beyond the dispersive strong regime. Our work provides a full quantitative description of the contribution to the qubit decoherence rate coming from the control line in chips that are currently employed in circuit QED laboratories, and suggests different possible ways to reduce this source of noise.

\end{abstract}

%% file: introduction.tex
\section{Introduction}

The study of the influence of environmentally induced noise on solid-state qubits is an active field both theoretically and experimentally \cite{jin-pra-2015,vespalainen-nature-2020,gordon-apl-2022,corcoles-apl-2011,wang-prl-2021}. A practical aim of understanding noise in quantum systems is to mitigate its effect on gate-based computing devices \cite{Krantz_quide}. In fact, quantum noise is directly relevant to applications in optimal quantum state control \cite{Kapit_2017}.

A common scenario in experimental implementations is that a quantum system ---comprised of a superconducting qubit and its readout resonator--- is thermalized to the base temperature of a dilution refrigerator measurement system with thermodynamic temperature of order $T=10\,\mathrm{mK}$. Despite this low temperature compared to relevant energy scales, the system must be connected to a remote bath of a higher temperature via control lines that enable qubit-specific measurement \cite{DiVincenzo-FdP-2000}.  Input signals are delivered to the device through control wires that are thermalized through a series of cryogenic microwave attenuators \cite{Krinner-epjqt-2019}.  An attenuator can act as a thermal reservoir that emits photons into the control line at the frequency of the readout resonator to induce random photon number fluctuations and dephasing via the a.c.~Stark effect for a qubit that is readout dispersively \cite{sears-prb-2012,gambetta-pra-2006}. Although design \cite{Yeh_2017,wang-prapp-2019} and placement \cite{Krinner-epjqt-2019} of attenuators can be optimized to decrease the temperature of the remote bath that is connected to a quantum system, it is not uncommon that photon shot-noise induces dephasing at rates that are much higher than what would be predicted from the base temperature of the device \cite{Wang_2019}. This effect motivates a detailed understanding of the interplay between a remote bath and quantum system for a dispersively coupled qubit from the starting point of its electrical circuit parameters. 

To distinguish between various functions of the solid-state qubit wiring in a cryogenic measurement system, the lines are typically classified as drive lines, flux lines, or output lines \cite{Krinner-epjqt-2019,Krantz_quide}.  The drive lines enable different functionalities dependent upon system architecture including: (1) delivery of signals to a readout resonator to detect its state, or (2) delivery of signals directly to a qubit through capacitive coupling for control pulses. In some superconducting qubit device designs (cf.~Ref.~\citenum{Burnett2019}), the same drive line can be used for both functions using frequency-domain multiplexing, and in other cases dedicated XY-control lines are utilized \cite{Barends_2013}.  However, the primary focus of this article is on the former type of drive line that is coupled to the readout resonator and throughout the article we use the terms "drive line" and "control line" interchangeably.

The scope of the present theoretical study is as follows.  We seek an accurate and complete solution to the dynamics of a superconducting qubit when it is in the presence of a remote thermal bath that is connected to the qubit--resonator system through its control lines. We obtain the solution through applying the theory of open quantum systems to the quantized circuit Hamiltonian of the Jaynes-Cummings form \cite{Shore_1993, Larson_2021}, which in turn is derived by applying the tools of circuit quantum electrodynamics. The resulting master equation is solved by exploiting its underlying symmetries \cite{Albert_2014}, allowing it to be transformed into a block diagonal form. This form simplifies our study of decoherence effects as we can focus only on one block of the whole Liouvillian matrix.

We study the decoherence rate by exploring broad parameter ranges of the values of resistor temperature, of qubit--resonator and of resonator--resistor couplings, and we focus in particular on both the dispersive strong and dispersive weak regimes \cite{Schuster_2007}. We notice that in the dispersive strong limit our model agrees with the often used approximation of the qubit decoherence rate being linearly proportional to the number of thermal photons in the readout resonator. However, in the dispersive weak regime the model predicts remarkably better decoherence rates as the dissipation rate of the resonator is increased. This suggests that working in the dispersive weak regime may significantly improve the decoherence time of the transmon qubit. 

The article is organized as follows. First we describe the derivation of the Hamiltonian and master equation in Sec.~\ref{sec:derivation}. We then present results of the model and discuss the solutions as they apply to experimentally relevant qubit and resonator parameters in the dispersive regime in Sec.~\ref{sec:solving}.  Within Sec.~\ref{subsec:direct_coupling} we briefly review the influence of a control line with direct capacitive coupling to the qubit. In Sec.~\ref{sec:concusions} we summarize the primary conclusions of the study and outlook for its impact on qubit circuit design.

%% file: model_derivation.tex
\section{Derivation of the circuit Hamiltonian and master equation} \label{sec:derivation}

In order to characterize the time evolution of the dispersively coupled transmon qubit \cite{Koch2007} we need to derive a master equation describing the dynamics of the quantum state. To obtain the Lindblad master equation, we first need to find a Hamiltonian describing the quantum system of interest, the environment, and the interactions \cite{Vaaranta_thesis}. The quantum system we consider is the superconducting circuitry of the qubit and its readout resonator, and the environment is an attenuator in the drive line, which is modelled as a bosonic bath. 

\begin{figure}
  \centering
  \resizebox{\linewidth}{!}{
  \begin{circuitikz}[cute inductors]
    \tikzstyle{every node}=[font=\Large]
    
    \draw (0,0) node[ground] {} to[R=$R$] (3,0) node[above] {$\dot{\phi}_\text{L}$} to[L, l_=$L_\text{L}$] (3,-3) node[ground] {};
    \filldraw (3,0) circle (2pt);

    \draw[<->, thick] (3.05,-1) to[in=135, out=45] node[above] {$M$} (3.95,-1);

    \draw (5,-0.5) -- (4,-0.5) to[L=$L_\text{f}$] (4,-2.5) -- (5,-2.5);
    \draw (5,0) to[C, l=$C_\text{g}$] (10,0);
    \draw (5,0) node[above] {$\dot{\phi}_\text{f}$} -- (5,-0.5) -- (6,-0.5) to[C,l_=$C_\text{f}$] (6,-2.5) -- (5,-2.5) -- (5,-3) node[ground] {};
    \filldraw (5,0) circle (2pt);

    \draw (10,0) node[above] {$\dot{\phi}_\text{A}$} -- (10,-0.5) -- (9,-0.5) -- (9,-1) -- (8.5,-1) -- node[left] {$V(\phi_\text{A})$} (8.5,-2) -- (9.5,-2) -- (9.5,-1) -- (9,-1);
    \draw (8.5,-1) -- (9.5,-2) -- (9.5,-1) -- (8.5,-2) -- (9,-2) -- (9,-2.5) -- (10,-2.5);
    \draw (10,-0.5) -- (11,-0.5) to[C=$C_\text{A}$] (11,-2.5) -- (10,-2.5) -- (10,-3) node[ground] {};
    \filldraw (10,0) circle (2pt);
    
  \end{circuitikz}
  }
  \caption{Lumped element circuit model. On the left is the drive line with inductive coupling to the readout resonator, which in turn is capacitively coupled to the qubit.}
  \label{fig:circuit-model}
\end{figure}
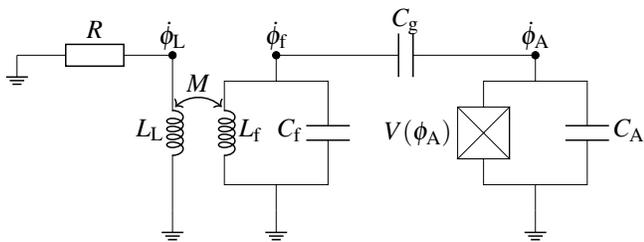

\subsection{Circuit Hamiltonian}

Our approach for deriving the Hamiltonian starts from the circuit diagram describing the qubit including its readout circuitry and the drive line. The form of the circuit we consider is motivated by the device in Ref.~\citenum{Burnett2019} with the inclusion of a resistive element in the drive line such as an attenuator. The circuit diagram is shown in Fig.~\ref{fig:circuit-model}, where the qubit is capacitively coupled to the readout resonator via capacitance $C_\text{g}$ and the resonator is inductively coupled to the drive line via mutual inductance $M$. It should be noted that in this model both the drive and readout signals travel through the readout resonator in the middle \cite{Burnett2019}.

As a physical implementation, the readout resonator is manufactured as a coplanar waveguide (CPW), which is a transmission line resonator with distributed inductance and capacitance values \cite{Goppl_2008}. We construct the lumped element circuit diagram in Fig.~\ref{fig:circuit-model} by treating the CPW resonator as a parallel LC-oscillator with fixed capacitance $C_\text{f}$ and inductance $L_\text{f}$. The exact values for these parameters can be derived from the CPW geometry \cite{Goppl_2008}. The transmission line readout resonators commonly used in circuit QED systems have many frequency harmonics. Approximating the resonator as a single LC mode as shown in Fig.~\ref{fig:circuit-model} assumes that the qubit transition frequency is far red-detuned from the fundamental mode of the resonator so that interaction with higher modes can be neglected in description of phenomena such as Purcell decay \cite{Houck-prl-2008}.  For this reason, using the single resonator mode $\omega_\text{f}$ is instructive and practical for the dispersive regimes described in the results of Sec.~\ref{sec:solving}.

Our goal is to derive the Hamiltonian of the circuit in Fig.~\ref{fig:circuit-model}. We first consider each energy term in the circuit to derive the classical Lagrangian, then move to the Hamiltonian description via the Legendre transformation, and finally quantize the classical Hamiltonian \cite{Vool_Devoret2017,Girvin2014}. The details of the derivation are discussed further in Appendix~\ref{appendix2} and Appendix \ref{appendix1}.

The general idea of the derivation is based on the representation of the resistor in the input line of Fig.~\ref{fig:circuit-model} as an infinite collection of harmonic oscillators (bosonic modes) with dense frequencies, which emit Johnson-Nyquist thermal noise \cite{Johnson1928,Nyquist1928a} through the input line. This sort of Caldeira-Leggett \cite{Caldeira-Leggett} representation of the resistor was introduced by Devoret in 1995 \cite{Devoret1995} and allows us to consider the physical degrees of freedom of the resistor as a circuital element as the dynamical variables (quadratures $q$ and $p$) of the collection of harmonic oscillators. In this way, we are able to introduce dissipation in the mathematical language of circuit QED by adding an infinite number of dynamical variables, which describe the modes of the resistor, to the circuit Hamiltonian. We refer the interested reader to Appendix~\ref{appendix2} and to Refs.~\citenum{Devoret1995,Cattaneo2021} for further details and explanations.

Our approach follows the standard mathematical description of quantum superconducting circuits \cite{Vool_Devoret2017,Girvin2014}. Within this framework, the dynamical variables that describe the circuit of interest are the node fluxes in the different nodes of the circuit in Fig.~\ref{fig:circuit-model}, and an additional set of node fluxes that are internal to the resistor. The time derivatives of these node fluxes give the voltage differences across each circuit element \cite{Vool_Devoret2017}. Finding the Hamiltonian of this dynamical system starting from the Lagrangian is not a trivial task, due to an infinite number of dynamical variables. To obtain the final solution, we rely on a method developed by Parra-Rodriguez et al. \cite{Parra_Rodriguez_2018,Parra-Rodriguez2021} and recently discussed in Ref.~\citenum{Cattaneo2021} from the perspective of open quantum systems. This method (see Appendix~\ref{appendix1} for details) makes use of a point transformation of the node fluxes to obtain a new set of dynamical variables that we label as $z$, and a set of conjugate momenta that we label as $p$. The corresponding circuit Hamiltonian is
\begin{equation}
  \label{eq:circuit-Hamiltonian}
  \begin{aligned}
    H &= \underbrace{\frac{p_\text{A}^2}{2}\frac{C_\text{f} + C_\text{g}}{D} + V(z_\text{A})}_{\text{Qubit}} + \underbrace{p_\text{A}p_\text{f}\frac{C_\text{g}}{D}}_{\substack{\text{Qubit-resonator} \\ \text{interaction}}} \\
    &+ \underbrace{\frac{p_\text{f}^2}{2}\frac{C_\text{A} + C_\text{g}}{D} + \frac{z_\text{f}^2}{2}\frac{1}{L_\text{f}}}_{\text{Resonator}}  + \underbrace{\sum_{k=1}^\infty\frac{M}{L_\text{L}L_\text{f}}f_k z_k z_\text{f}}_{\substack{\text{Resistor-resonator} \\ \text{interaction}}} \\
    &+ \underbrace{\sum_{k=1}^\infty\Bigg(\frac{p_k^2}{2M_0}\frac{1}{L_k C_k} + \frac{M_0}{2}z_k^2\Bigg)}_{\text{Resistor}}\,.
  \end{aligned}
\end{equation}
In Eq.~\eqref{eq:circuit-Hamiltonian} the variables $z_\text{A}$, $z_\text{f}$ and $z_k$ describe the transformed fluxes of the qubit, resonator and resistor nodes respectively. The variables $p_\text{A}$, $p_\text{f}$ and $p_k$ are the corresponding conjugate momenta. The coefficient $D = C_\text{A}C_\text{f} + C_\text{f}C_\text{g} + C_\text{g}C_\text{A}$ is the determinant of the upper-left corner block of the capacitance matrix $\bm{C}$ \cite{Girvin2014} describing the qubit--resonator system, and $M_0$ is some constant with units of inverse inductance (see Appendix~\ref{appendix1} for details). $V(z_\text{A})$ denotes nonlinear inductive potential of the qubit, that will give rise to the anharmonicity of the qubit energy levels.

We observe that the Hamiltonian in Eq.~\eqref{eq:circuit-Hamiltonian} can be written as $H = H_\text{System} + H_\text{Bath} + H_\text{Int}$, which is the standard structure of the Hamiltonian of an open quantum system \cite{BreuerPetruccione}. Here the system part includes the terms for the qubit, the resonator and their interaction. The bath term includes only the resistor contribution, while $H_\text{Int}$ describes the interaction between the (qubit--resonator) system and the bath (the resistor).

Eq.~\eqref{eq:circuit-Hamiltonian} behaves well in the limits of vanishing qubit--resonator coupling ($C_\text{g} \to 0$) and vanishing resistor--resonator coupling ($M \to 0$). We can see that in these limiting cases the corresponding terms of the Hamiltonian decouple as the interactions vanish.

In order to be able to  quantize the Hamiltonian Eq.~ \eqref{eq:circuit-Hamiltonian} completely, we need to set the inductive potential of the qubit $V(z_\text{A})$ explicitly. Here we use the transmon potential \cite{Koch2007} $V(z_\text{A}) = -E_\text{J}\cos\Big(2\pi z_\text{A}/\phi_0\Big)\,,$ where $E_\text{J}$ is the Josephson energy and $\phi_0$ the magnetic flux quantum. We assume now that the manufactured qubit is in the transmon regime characterized by low anharmonicity ($E_J\gg E_C$ \cite{Koch2007}) and, simultaneously, that it is operated at a thermodynamic temperature sufficiently low so that we can neglect the excitations to second and higher energy levels \footnote{Note that this is not in contradiction with the presence of a hot resistor in the input line. In fact, we will assume to be in the dispersive regime of the qubit--resonator coupling. Therefore, the thermal fluctuations coming from the input line will affect the qubit coherences only, while they will not induce any changes in the energy of the qubit, i.e., they will not create excitations in the higher qubit levels.}. Thus we can confidently truncate the Hilbert space of the transmon qubit to the ground state $\ket{g}$ and first excited state $\ket{e}$ only.

After the quantization procedure discussed in more detail in the Appendix \ref{appendix1}) the circuit Hamiltonian can be written as,
\begin{equation}
  \label{eq:Quantized-Circuit-Hamiltonian-final}
  \begin{aligned}
  H &= \underbrace{\frac{1}{2}\hbar\omega_\text{A}\sigma_z}_{\text{Qubit}} + \underbrace{\hbar\omega_\text{f}a^\dagger_\text{f}a_\text{f}}_{\text{Resonator}} + \underbrace{\sum_{k=1}^\infty\hbar\omega_k a_k^\dagger a_k}_{\text{Resistor}} \\
  &- \underbrace{\i\hbar g_\text{f}\sigma_y(a^\dagger_\text{f} - a_\text{f})}_{\substack{\text{Qubit-resonator} \\ \text{interaction}}}  + \underbrace{\frac{M}{L_\text{L}}\sum_{k=1}^\infty h_k (a^\dagger_k + a_k)(a^\dagger_\text{f} + a_\text{f})}_{\text{Resistor-resonator interaction}}\,.
  \end{aligned}
\end{equation}
The Pauli matrices act on the truncated Hilbert space of the qubit. The ladder operators $a_\text{f}^\dagger/a_\text{f}$ and $a_k^\dagger/a_k$ create and destroy excitations respectively in the single mode of the resonator and in the $k$\textsuperscript{th} mode of the resistor. The coefficient $h_k$ describes the interaction strength between the resonator and the mode $k$ of the resistor. Its form is given in Eq.~\eqref{eq:g_k} of Appendix~\ref{appendix1}. The qubit--resonator coupling strength $g_\text{f}$ is given in terms of the circuit parameters from the circuit in Fig.~\ref{fig:circuit-model} as
\begin{equation}
  \label{eq:gf}
  g_\text{f} \equiv \frac{C_\text{g}}{2}\sqrt{\frac{\omega_\text{A}\omega_\text{f}}{(C_\text{f} + C_\text{g})(C_\text{A} + C_\text{g})}}\,.
\end{equation}

Before discussing the derivation of the master equation, we can compute the spectral density of the bath by using its standard definition \cite{BreuerPetruccione,Cattaneo2021}. We obtain
\begin{align}
  \label{eq:spectral-density}
  J(\omega) &= \frac{1}{\hbar^2}\sum_{k=1}^\infty|h_k|^2\delta(\omega-\omega_k) \nonumber \\ 
  &= \chi\frac{\omega_\text{c}^2}{\pi\omega(\omega_\text{c}^2+\omega^2)}\,,
\end{align}
where we introduced a factor $\chi = R\mu^2/(\hbar L_\text{f}^2)$, where $\mu$ can be found in Eq.~\eqref{eq:mu} of Appendix~\ref{appendix1}.

\subsection{The master equation}

 For the microscopic derivation of the master equation, we need to find so called "jump operators" of the system \cite{BreuerPetruccione, lidar2020lecture, Cattaneo_2019}, which requires diagonalizing the system Hamiltonian $H_\text{Sys} = 1/2\hbar\omega_\text{A}\sigma_z + \hbar\omega_\text{f}a^\dagger_\text{f}a_\text{f} - \i\hbar g_\text{f}\sigma_y(a^\dagger_\text{f} - a_\text{f})$. We do this by moving to the dispersive frame \cite{Blais2004}. In the dispersive frame we assume that the detuning between the qubit and resonator resonance frequencies, $\Delta = |\omega_\text{A} - \omega_\text{f}|$, is large compared to the coupling strength $g_\text{f}$ between them, such that $\lambda = g_\text{f}/\Delta \ll 1$. This process, which is shown in detail in Appendix \ref{appendix3}, will give us the dispersive Jaynes-Cummings Hamiltonian \cite{Boissonneault_2009} with added interaction between the resonator and the resistor embedded in the qubit drive line,
\begin{equation}
  \label{eq:Disperisve-Hamiltonian-in-pieces}
  H^\text{D} = H_\text{Sys}^\text{D} + H_\text{Bath} + H_\text{Int}^\text{D}\,,
\end{equation}
where
\begin{align}
\begin{split}
  H_\text{Sys}^\text{D} = {}&\frac{1}{2}\hbar(\omega_\text{A} + 2 g_\text{f}\lambda)\sigma_z \\
  &+ \hbar(\omega_\text{f} + g_\text{f}\lambda\sigma_z)a_\text{f}^\dagger a_\text{f} + \hbar g_\text{f}\lambda\sigma_-\sigma_+\,,
  \end{split}\label{eq:H-D-system}\\
  H_\text{Bath} = &\sum_{k=1}^\infty\hbar\omega_k a_k^\dagger a_k\,, \label{eq:H-D-bath} \\
  H_\text{Int}^\text{D} = &\frac{M}{L_\text{L}}(a^\dagger_\text{f} + a_\text{f} + \lambda\sigma_x)\sum_{k=1}^\infty h_k (a^\dagger_k + a_k) \label{eq:H-D-interaction}\,.
\end{align}
Here the superscript $^\text{D}$ denotes the dispersive frame. We can see that in this frame the qubit--resonator part given in Eq.~\eqref{eq:H-D-system} is diagonal in the $\{\ket{e,\ n-1}, \ket{g,\ n}\}$ basis, where the first part denotes the state of the qubit and the second part denotes the number of quanta in the resonator. The eigenenergies of the different eigenstates are given by
\begin{equation}
  \label{eq:eigenenergies}
  \begin{aligned}
    E_{g,n} &= \Big(n-\frac{1}{2}\Big)\hbar\omega_\text{f} - \frac{\hbar\Delta}{2} - n\hbar g_\text{f}\lambda\,, \\
    E_{e,n-1} &= \Big(n-\frac{1}{2}\Big)\hbar\omega_\text{f} + \frac{\hbar\Delta}{2} + n\hbar g_\text{f}\lambda \,.\\
  \end{aligned}
\end{equation}

The dispersive Hamiltonian from Eq.~\eqref{eq:Disperisve-Hamiltonian-in-pieces} along with the eigenenergies in Eq.~\eqref{eq:eigenenergies} are the starting points for the derivation of the master equation, which describes the dynamics of the qubit--resonator system. The derivation of the master equation along with a careful consideration of the secular approximation \cite{Cattaneo_2019} is discussed in Appendix \ref{appendix4}. To obtain the master equation, we make additional use of two important approximations: i) Born-Markov approximation, which requires weak coupling between resonator and resistor ($M/L_L\ll \omega_\text{f}/\omega_\text{A}$ in Eq.~\eqref{eq:Quantized-Circuit-Hamiltonian-final}, which sets the resonator-resistor interaction energy to be much smaller than the energy of the resonator) and that the autocorrelation functions of the bath decay sufficiently fast in time \cite{BreuerPetruccione}. The latter is typically true for the Johnson-Nyquist thermal fluctuations of the voltage across the resistor \cite{Cattaneo2021}. ii) Approximation of local master equation \cite{Trushechkin2016,Cattaneo_2019,Hofer2017,Scali2020}, valid when the coupling energy between the transmon qubit and the resonator is much smaller than their energies ($g_{\text{f}}\ll\omega_\text{A},\omega_\text{f}$); as we are in the dispersive regime, this approximation is always valid. Finally, the master equation we obtain is the following:
\begin{equation}
  \label{eq:ME!!}
  \begin{aligned}
    \dot{\rho}_\text{Sys} =\mathcal{L}[\rho_\text{Sys}]= &-\frac{\i}{\hbar}\big[H_\text{Sys}^\text{D} + H_{\text{LS}}, \rho_\text{Sys}\big] \\
    &+ \gamma\bar{n}\Big(a^\dagger_\text{f}{\rho}_\text{Sys}a_\text{f} - \frac{1}{2}\big\{a_\text{f}a^\dagger_\text{f}, {\rho}_\text{Sys}\big\}\Big) \\
    &+ \gamma(1+\bar{n})\Big(a_\text{f}{\rho}_\text{Sys}a^\dagger_\text{f} - \frac{1}{2}\big\{a^\dagger_\text{f}a_\text{f}, {\rho}_\text{Sys}\big\}\Big)\,.
  \end{aligned}
\end{equation}
$\mathcal{L}$ is the generator of the Markovian master equation driving the system dynamics, and it is the so-called ``Liouvillian superoperator'' \cite{BreuerPetruccione,Cattaneo_2020}. The Lamb-shift Hamiltonian $H_\text{LS}$ arises from the environmental interaction and effectively renormalizes the resonator's resonance frequency. $H_\text{LS}$ does not have a considerable effect on the qubit dynamics, therefore we will neglect it in the following discussion. The coefficient $\gamma$ is the dissipation rate of the resonator and it is dependent on the spectral density of the bath as
\begin{equation}
    \label{eq:gamma_decay_rate}
    \gamma = 2\pi\alpha^2J(\omega_f),
\end{equation} with $\alpha=M/L_\text{L}$ describing the interaction strength between the bath and the resonator. The factor $\bar{n}$ is the expected number of quanta in the resonator. Following the Bose-Einstein distribution, it depends on temperature $T$ and the resonance frequency of the resonator $\omega_\text{f}$ \cite{BreuerPetruccione}.

%% file: solving_models.tex
\section{Results and discussion} \label{sec:solving}

Now that we have obtained the master equation describing the time evolution of the dispersively coupled transmon coupled to the resistive bath in the drive line, we need to solve it for the qubit dynamics. The master equation of the type in Eq.~\eqref{eq:ME!!} has been solved analytically at the absolute zero \cite{Peixoto-de-Faria_1999} and more generally in Ref.~\citenum{Obada_2008}, both studies assuming that the atom (corresponding to our resonator) is in a coherent state. Our approach to solving the master equation involves finding out its inherent symmetries and representing the Liouvillian superoperator as a block diagonal matrix \cite{Cattaneo_2020}. This approach reveals an analytical expression for the long-time decoherence rate of the qubit and makes the numerical solution of the dynamics easier by reducing the number of needed Liouvillian matrix elements from $s^4S^4$ to $s^2S^2$, where $s$ is the truncation of the qubit's Hilbert space and $S$ is the size of Hilbert space truncation for the resonator. Moreover our solution makes no \textit{a priori} assumptions about the state of the system but can be utilized to compute the dynamics of an arbitrary initial state. The block structure of the Liouvillian is discussed in more detail in appendix \ref{appendix5}.

Before considering the dynamics of the dispersive transmon with decoherence, we review the solution for a simple XY-control of the qubit where the resistor in the drive line is directly coupled to the transmon. In this simpler case the solution is obtainable analytically. Comparing this scenario with the one with the readout resonator in between shows that the addition of the resonator in the dispersive regime yields pure dephasing and no dissipation in the qubit dynamics.

\subsection{Dissipation in the case of direct resistor--qubit coupling \label{subsec:direct_coupling}}

A drive line directly coupled to the qubit without any intermediate resonator in between is used for the XY-control of the qubit \cite{Barends_2013} as it couples the qubit's $\sigma_x$ and $\sigma_y$ operators to the bath operators, allowing for the excitation of the qubit energy levels. The effect of coupling a resistor directly to the qubit is discussed in detail in Ref.~\citenum{Cattaneo2021}. This direct coupling is described by the following Hamiltonian:
\begin{equation}
    H = \frac{1}{2}\hbar\omega_\text{A}\sigma_z + \sum_{k=1}^\infty\hbar\omega_k a^\dagger_k a_k + \zeta\sum_{k=1}^\infty g_k\sigma_y(a_k^\dagger - a_k)\,,
\end{equation}
where $\zeta$ describes the interaction strength between the qubit and the resistor. In this case we can notice that the qubit experiences transverse noise through the $\sigma_y$ coupling, which leads to dissipation. Taking the qubit--resistor coupling to be weak, the above Hamiltonian leads into a master equation describing the state of the qubit $\rho_\text{Q}$ at some time $t$,
\begin{equation}
  \begin{aligned}
    \dot{\rho}_\text{Q} = &-\frac{\i}{\hbar}\Big[\frac{1}{2}\hbar\omega_\text{A}\sigma_z + H_{\text{LS}}, \rho_\text{Q}\Big] \\
    &+ \Gamma_\downarrow\Big(\sigma_-\rho_\text{Q}\sigma_+ - \frac{1}{2}\big\{\sigma_+\sigma_-, {\rho}_\text{Q}\big\}\Big) \\
    &+ \Gamma_\uparrow\Big(\sigma_+\rho_\text{Q}\sigma_- - \frac{1}{2}\big\{\sigma_-\sigma_+, {\rho}_\text{Q}\big\}\Big)\,.
  \end{aligned}
\end{equation}
Here the absorption and emission rates $\Gamma_\uparrow$ and $\Gamma_\downarrow$ depend on the bath temperature and its spectral density, which is proportional to the number of thermal photons in the resistor, given by the Bose-Einstein distribution. This master equation leads to decay of qubit populations, that is dissipation, given by the rate $\Gamma_\uparrow + \Gamma_\downarrow$, and decay of the coherences, that is decoherence, given by the rate $(\Gamma_\uparrow + \Gamma_\downarrow)/2$ \cite{BreuerPetruccione}. The rate of dissipation and decoherence grows linearly as the number of thermal photons in the resistor increases with temperature.

\subsection{Decoherence through dispersive qubit--readout resonator coupling} \label{decoher for disp transmon}

Let us now consider the case with a resonator between the qubit and the resistor, which is the central focus of this work.
The master equation of Eq.~\eqref{eq:ME!!} describes the decay of a resonator that is dispersively coupled to the transmon qubit. So, we may expect to observe no dissipation on the quantum state of the qubit. This can be proved by computing the adjoint master equation, which is essentially the Heisenberg picture equivalent to the normal Schrödinger picture master equation \cite{BreuerPetruccione}. In the adjoint master equation the time dependency is transferred from the states to the operators. In our case the adjoint master equation is
\begin{equation}
  \label{eq:adjoint-ME}
  \begin{aligned}
    \dot{O}(t) = \mathcal{L}^\dagger[O(t)]=&\frac{\i}{\hbar}\big[H^\text{D}_\text{Sys}, O(t)\big] \\
    &+ \gamma\bar{n} \Big(a_\text{f}O(t)a^\dagger_\text{f} - \frac{1}{2}\big\{a_\text{f}a^\dagger_\text{f}, \, O(t)\big\} \Big) \\
    &+ \gamma(1+\bar{n}) \Big(a^\dagger_\text{f}O(t)a_\text{f} - \frac{1}{2}\big\{a^\dagger_\text{f}a_\text{f}, \, O(t)\big\} \Big)\,.
  \end{aligned}
\end{equation}
 Setting $O = \sigma_z$ in the above equation will yield us the time evolution of the $\sigma_z$ operator describing the qubit populations. It is easy to see that $\dot{\sigma}_z = 0$ because $\sigma_z$ commutes with the ladder operators $a_\text{f}$ and $a_\text{f}^\dagger$, which sets the dissipator (i.e., the terms in the master equation proportional to $\gamma$ that are responsible for the non-unitary open dynamics) to zero. It also commutes with the dispersive system Hamiltonian in Eq.~\eqref{eq:H-D-system} and therefore also the unitary part of the master equation vanishes. Thus $\dot{\sigma}_z = 0$ and so the qubit experiences no dissipation.

One key issue in the theory of open quantum systems is the search for the steady states of the dynamics \cite{BreuerPetruccione}. Since, as stated above, total energy is a conserved quantity in the system dynamics, the steady state is not unique \cite{Albert_2014}. The subspace of the steady states is two-dimensional and it is spanned by product states of the qubit and resonator. The qubit part is easy to find to be either $\ket{g}\bra{g}$ or $\ket{e}\bra{e}$. The resonator contribution turns out to be the thermal state given by 
\begin{equation}
    \rho_{\text{Th}}(\omega) = \frac{1}{1+\bar{n}(\omega)}\sum_{n=0}^\infty\Big(\frac{\bar{n}(\omega)}{1 + \bar{n}(\omega)}\Big)^n\ket{n}\bra{n}\,,
    \label{thermal state}
\end{equation}
where $\bar{n}(\omega)$ is the Bose-Einstein distribution. So the most general steady state of the system can be written as
\begin{equation}
  \label{eq:steady-state}
  \rho_\text{Steady} = a\ket{e}\bra{e}\otimes\rho_\text{Th} + (1-a)\ket{g}\bra{g}\otimes\rho_\text{Th} \,,
\end{equation}
where $a\in[0,1]$. If there are no oscillating coherences \cite{Albert_2014} (i.e., coherences that survive and oscillate even at infinite time), this is the family of steady states towards which we can expect the combined qubit--resonator quantum state to evolve in time. We will later see that we can find oscillating coherences only in the case of zero temperature or very small ratio $\gamma^\prime/g_\text{f}^{\prime 2}$, where the primed parameters are normalized with the qubit frequency $\omega_\text{A}$.

We solve the master equation \eqref{eq:ME!!} by transforming it into the Liouville space \cite{Gyamfi_2020}, where the quantum state density operators become vectors and the Liouvillian superoperators are represented as matrices operating on the state vectors. In this formalism the Liouvillian superoperator describing the master equation can be written as
\begin{equation}
  \label{eq:Liouvillian-matrixified}
  \begin{aligned}
    \mathcal{L} = &-\frac{\i}{\hbar}\big(H^\text{D}_\text{Sys}\otimes\mathbb{1} - \mathbb{1}\otimes H^{\text{D}\top}_\text{Sys}\big) \\
    &+ \gamma\bar{n}\Big(a_\text{f}^\dagger\otimes a_\text{f}^\dagger - \frac{1}{2}a_\text{f}a_\text{f}^\dagger\otimes\mathbb{1} - \frac{1}{2}\mathbb{1}\otimes a_\text{f}a_\text{f}^\dagger\Big) \\
    &+ \gamma(1+\bar{n})\Big(a_\text{f}\otimes a_\text{f} - \frac{1}{2}a_\text{f}^\dagger a_\text{f}\otimes\mathbb{1} - \frac{1}{2}\mathbb{1}\otimes a_\text{f}^\dagger a_\text{f}\Big)\,.
  \end{aligned}
\end{equation}
It can be shown (see Appendix~\ref{appendix5}) that $\mathcal{L}$ as a superoperator matrix commutes with the total-number-of-particles superoperator defined as $\mathcal{N}=N\otimes\mathbb{1} - \mathbb{1}\otimes N^\top$ in Liouville space, with $N=\sigma_z +  a_\text{f}^\dagger a_\text{f}$. Therefore, we can block-diagonalize $\mathcal{L}$ by labeling each block through an eigenvalue of $\mathcal{N}$. It turns out that to study the evolution of the qubit coherences (i.e., the mean values of $\sigma_x(t)$ and $\sigma_y(t)$) we can focus on a single Liouvillian block only, namely $\mathcal{L}_1$, associated with the eigenvalue $1$ of $\mathcal{N}$. Therefore, the property $[\mathcal{L},\mathcal{N}]=0$ simplifies the solution of the master equation for the observables we need to track to study the qubit decoherence. The dimension of the block $\mathcal{L}_1$ is in principle infinite, but will depend on the truncation of the Hilbert space of the resonator when computing the dynamics numerically. The block diagonalization of the Liouvillian is discussed in detail in Appendix \ref{appendix5}.

To understand how the thermal noise coming from the resistor affects the qubit coherences, we study the behavior of a coherence measure, such as the $l_1$-norm \cite{Baumgratz_2014}:
\begin{equation}
  \label{eq:coherence-measure}
  C(\rho) = \sum_{\substack{i,j \\ i\neq j}}|\rho_{ij}|\,.
\end{equation}
In the two dimensional case of the qubit Hilbert space the coherence measure is simply given by,
\begin{equation}
  \label{eq:coherence-measure-2}
  C(\rho_\text{Q}) = \sum_{\substack{i,j \\ i\neq j}}|\rho_{ij}| = |\rho_{01}| + |\rho_{10}| = \sqrt{\langle\sigma_x\rangle^2 + \langle\sigma_y\rangle^2}\,.
\end{equation}
The required expectation values $\langle\sigma_x\rangle$ and $\langle\sigma_y\rangle$ can be obtained from the eigenmodes of the Liouvillian block $\mathcal{L}_1$ as \cite{Bellomo_2017, Cattaneo_2021_bath}
\begin{equation}
  \label{eq:expectation-of-sigmax3}
  \langle\sigma_x(t)\rangle = \sum_i2|c_i^{(1)}p_i|\e^{\text{Re}[\lambda_i^{(1)}] t}\cos\big(\text{Im}[\lambda_i^{(1)}]t + \text{Arg}[c_i^{(1)}p_i]\big)\,.
\end{equation}
Here the coefficients $c_i^{(1)}$ describe the initial conditions of the qubit--resonator quantum state, $p_i$ describe the projection of the vectorized operator $\sigma_x$ into the eigenvectors of the Liouvillian and $\lambda_i^{(1)}$ are the eigenvalues of the Liouvillian block with $d=1$, where $d$ is an eigenvalue of $\mathcal{N}$ (see Appendix \ref{appendix5} for details). The expectation value of $\sigma_y$ are computed in a similar manner but with $p_i$ describing the projection of $\sigma_y$ onto the suitable eigenspaces of $\mathcal{L}_1$. Importantly, Eq.~\eqref{eq:expectation-of-sigmax3} reveals that the coherences can be thought of as being constructed from a sum of different modes of the Liouvillian, each with a specific eigenvalue $\lambda_i^{(1)}$ that determine how fast that mode decays in time due to the exponential factor $\e^{\text{Re}[\lambda_i^{(1)}] t}$. Note that all of the Liouvillian eigenvalues are non-positive, with the eigenvalue zero corresponding to the steady state \cite{Albert_2014}.

 To quantify the rate of decoherence due to the presence of the resistor, we define the thermalization rate (being the rate at which the qubit coherence decays in time) as the slowest decaying mode of the Liouvillian block $d=1$ given by,
\begin{equation}
  \label{eq:T2-timescale}
  \Gamma_{2, \text{R}} = \min_i\big(\big|\text{Re}\big[\lambda_i^{(1)}\big]\big|\big)\,.
\end{equation}
The equation above expresses the inverse of the timescale for the complete disappearance of all the qubit coherence. However, this does not give us any information about the rate of qubit decoherence at any instant of the dynamics. For instance, we will see that the decoherence rate of the qubit at an arbitrary time $t$ also depends on the initial state of the resonator. This decoherence rate is slower (approaching Eq.~\eqref{eq:T2-timescale}) when the reduced state of the resonator gets closer to its thermal state with no off-diagonal elements (see Appendix \ref{appendix6} for further discussions and some numerical examples).

The eigenvalues $\lambda_i^{(1)}$ in Eq.~\eqref{eq:T2-timescale} depend on all of the circuit parameters used to derive the circuit Hamiltonian and the master equation. Therefore we are able to compute the qubit dynamics in terms of the physical circuit parameters. Moreover, the eigenvalues depend also on the temperature of the environment, so we can find the temperature dependency of the decoherence time by computing the eigenvalues $\lambda_i^{(1)}$ for different temperatures and applying Eq.~\eqref{eq:T2-timescale} to pick the one that describes the slowest decay. 

\subsection{Decoherence blockade at zero temperature} \label{sec:decoherenceBlockadeZeroT}

The master equation [Eq.~\eqref{eq:ME!!}] has a unitary part $\mathcal{U}$, which is the standard Liouville-von Neumann equation, and the dissipative part $\mathcal{D}$, which is responsible for the decoherence effects. In general, we can write the Liouvillian superoperator as $\mathcal{L} = \mathcal{U} + \mathcal{D}$ \cite{BreuerPetruccione,Cattaneo_2020}. Looking at this structure, we can readily make an important remark: if the dissipative superoperator $\mathcal{D}$ commutes with the unitary superoperator $\mathcal{U}= -\i/\hbar[H_\text{Sys}, \,\cdot]$ on the subspace of states written as $\rho_Q\otimes \rho_\text{Th}$, where $\rho_\text{Th}$ is the thermal state of the resonator, then there will be a subspace of the dynamics with no decoherence. In other words, if $\restr{[\mathcal{U},\mathcal{D}]}{\rho_Q\otimes \rho_\text{Th}}=0$, then $\Gamma_{2,R}$ in Eq.~\eqref{eq:T2-timescale} will be zero, i.e., there will be some \textit{oscillating coherences} \cite{Albert_2014} in the qubit dynamics. Indeed, if the unitary propagator and the dissipator commute on this subspace, then the action of the Liouvillian on the same subspace can be separated into two parts as 
\begin{align}
    \rho(t) &= \e^{\mathcal{L}t}\rho(0) = \e^{\mathcal{U}t}\e^{\mathcal{D}t}\rho(0),
\end{align}
for $\rho(0) = \rho_\text{Q}\otimes\rho_\text{Th}$. Using $\mathcal{D}[\rho_\text{Th}] = 0$, we get
\begin{equation}
\label{eq:unitaryEv}
    \rho(t) = \e^{\mathcal{U}t}[\rho_\text{Q}\otimes\rho_\text{Th}].
\end{equation}
That is to say, the qubit and the resonator follow a purely unitary evolution, therefore the qubit will never lose all of its coherences.

Let us now compute the commutator $[\mathcal{U},\mathcal{D}]$ for the master equation in Eq.~\eqref{eq:ME!!} on a separable state $\rho_\text{Q}\otimes\rho_\text{Th}$. We find:
\begin{equation}
     [\mathcal{U}, \mathcal{D}]\rho_\text{Q}\otimes\rho_\text{Th} =\i g_\text{f}\lambda[\sigma_z, \rho_\text{Q}]\mathcal{D}[a_\text{f}^\dagger a_\text{f}\rho_{\text{Th}}]\,.
     \label{eq:superoperator commutator}
\end{equation}

In the limit of zero temperature the thermal state of the resonator is simply $\ket{0}\bra{0}$, as can be easily observed from Eq.~\eqref{thermal state}. This sets the dissipator part $\mathcal{D}[a_\text{f}^\dagger a_\text{f}\ket{0}\bra{0}] = 0$, which thus leads to commuting unitary and dissipative parts. Therefore, at absolute zero the system does not experience any decoherence when the resonator is initialized in the ground state.
If the resonator is initially not in a thermal state, then there will be a transient dynamics during which the resonator reaches the thermal state and the qubit is allowed to decohere. After this transient, the joint state of system and resonator will once again be written as $\rho_\text{Q}\otimes\ket{0}\bra{0}$, and all the remaining qubit coherences will oscillate in a purely unitary fashion according to Eq.~\eqref{eq:unitaryEv}. Additionally, if the quantum state of the qubit commutes with the $\sigma_z$ operator, then no coherences are present and thus no decoherence can happen, which is also captured by Eq.~\eqref{eq:superoperator commutator}.

\subsection{Numerical solutions}

In order to solve the master equation numerically, we begin with the Liouvillian in Eq.~\eqref{eq:Liouvillian-matrixified}. The solution for some arbitrary time $t$ can be obtained by exponentiating the Liouvillian matrix: $\rho(t) = \e^{\liouv t}\rho(0)$. However, to use this method, the generally infinite dimensional Hilbert space of the ladder operators $a_\text{f}$ and $a_\text{f}^\dagger$ has to be truncated to a finite size $S$, which describes the dimensionality of the matrix representation of the ladder operators. The truncation $S$ effectively sets the upper limit for the resonator excitations. Therefore, we choose $S$ so that the resonator has a sufficient amount of available exited states, such that it can get close enough to its ideal steady state, the thermal state. In practice we choose $S$ such that the probability of occupying the highest energy state for a given temperature $T$ is $10^{-7}$ or less.

In order to compute the qubit dynamics of the circuit in Fig.~\ref{fig:circuit-model}, we set the values for the circuit elements to be the following: the resistance $R=50\,\mathrm{\Omega}$, the capacitances $C_\text{f}$, $C_\text{g}$ and $C_\text{A}$ are $800\,\mathrm{fF}$, $5\,\mathrm{fF}$ and $90\,\mathrm{fF}$ respectively and the resonance frequencies of the qubit and resonator are $\omega_\text{A} = 2\pi\cdot4\,\mathrm{GHz}$ and $\omega_\text{f} = 2\pi\cdot6.1\,\mathrm{GHz}$ respectively. The coupling inductance is $L_\text{L} = 140\,\mathrm{pH}$ and the mutual inductance is $M = k\sqrt{L_\text{L}L_\text{f}}$, with $k=0.005$ and $L_\text{f}$ is obtained from the resonator's capacitance and resonance frequency as $L_\text{f} = 1/(C_\text{f}\omega_\text{f}^2)$.

The values listed above determine the dynamics of the qubit as the qubit-resonator coupling $g_\text{f}$ and the environmental coupling parameter $\gamma$ are completely determined as a function of the circuit parameters (see Appendix~\ref{appendix1} and Eqs.~\eqref{eq:gf} and~\eqref{eq:spectral-density}). The coupling rates can be written as
\begin{equation}
  \label{eq:g_f}
  \begin{aligned}
  g_\text{f}/(2\pi) &= \frac{C_\text{g}}{2}\sqrt{\frac{\omega_\text{A}\omega_\text{f}}{(C_\text{f} + C_\text{g})(C_\text{A} + C_\text{g})}} \approx 44.7\,\mathrm{MHz} \\
  \Rightarrow \hspace{0.25cm} g_\text{f}^\prime &= \frac{g_\text{f}}{\omega_\text{A}} \approx 0.0112
  \end{aligned}
\end{equation}
and
\begin{multline}
  \label{eq:gamma}
  \gamma/(2\pi) = 2\pi\Bigg(\frac{M}{L_\text{L}}\Bigg)^2\frac{R(C_\text{A} + C_\text{g})}{2(C_\text{A}C_\text{f} + C_\text{f}C_\text{g} + C_\text{g}C_\text{A})\omega_\text{f} L_\text{f}^2} \\
  \times\frac{\omega_\text{c}^2}{\pi\omega_\text{f}(\omega_\text{c}^2+\omega_\text{f}^2)} \approx 1.41\,\mathrm{MHz} \nonumber
\end{multline}
\begin{equation}
\label{eq:gamma_prime}
    \Rightarrow \hspace{0.25cm} \gamma^\prime = \frac{\gamma}{\omega_\text{A}} \approx 3.53\cdot10^{-4}\,.
\end{equation}
The cutoff frequency $\omega_\text{c}$ is taken to be $1\,\mathrm{THz}$.

\begin{figure}
    \centering
    \includegraphics[width=\linewidth]{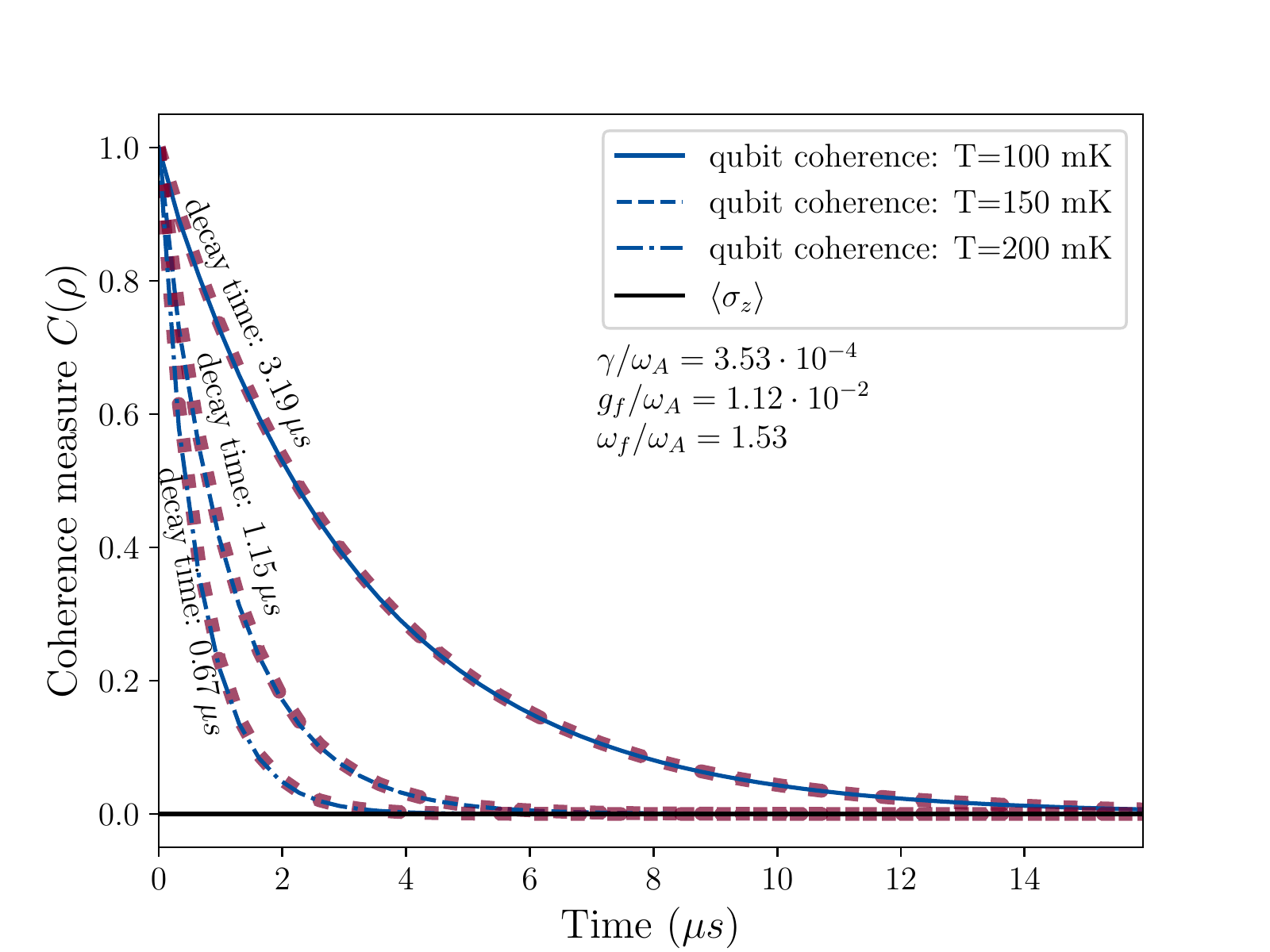}
    \caption{Time evolution of the qubit coherence measure $C(\rho) = \sqrt{\langle\sigma_x\rangle^2 + \langle\sigma_y\rangle^2}$ for different temperatures in blue. Red squares show pure decaying exponential with the decay time computed using Eq.~\eqref{eq:T2-timescale}. We observe that the qubit coherences decay almost exactly as predicted by this equation. The listed parameters have been normalized by the qubit frequency $\omega_\text{A}$.}
    \label{fig:real-qubit-coherence-decay}
\end{figure}

The master equation is numerically solved for the given values and the time evolution of the qubit coherence is presented in Fig.~\ref{fig:real-qubit-coherence-decay} for three different resistor temperatures. The qubit was initially in the superposition state $\ket{\Psi_\text{Q}} = \ket{+} = \frac{1}{\sqrt{2}}\big(\ket{g} + \ket{e}\big)$ and the resonator in the thermal state for the given temperature. The blue lines show numerically exact solutions to the qubit time evolution. We can see that they almost exactly match an exponential decay, marked by the red squares, where the decay rate was computed using the decoherence rate from the resistor in Eq.~\eqref{eq:T2-timescale}. This corroborates the validity of Eq.~\eqref{eq:T2-timescale} to estimate the long-time decoherence rate, and shows that the qubit experiences a slower decoherence when the reduced state of the resonator is its thermal state. This means that when the resonator is in a thermal state there is a decaying eigenmode of the Liouvillian that is the only relevant contribution to the system evolution, as can be deduced from Eq.~\eqref{eq:expectation-of-sigmax3}. In other cases where the resonator does not start in the thermal state, the other modes of the Liouvillian give a non-trivial contribution to the system dynamics, and decay at a faster rate until the resonator reaches its steady state, at which point the rest of the dynamics is once again mostly governed by the exponential decay rate given by Eq.~\eqref{eq:T2-timescale}. Further discussions can be found in Appendix \ref{appendix6}.

\subsection{Dependence of $T_2$ on resistor temperature}

The noise emanating from the qubit drive line resistor is not the only source of decoherence (see Ref.~\citenum{Krantz_quide} and references therein). Other factors contributing to the coherence loss are for example quasiparticle tunneling \cite{Catelani_2012}, magnetic flux noise of the field penetrating the circuit \cite{Ithier_2005} and phonon radiation of the surrounding lattice \cite{Ioffe_2004}. All of these factors have their own non-trivial contribution to the physical decoherence rate. These different contributions can, in general, depend on temperature and on other aspects of the circuit and the environment but for simplicity we assume their contribution to the true decoherence rate to be some constant amount $\Gamma_\text{B}$, that we term the background decay rate. This background decay rate together with the decoherence rate coming from the resistor $\Gamma_{2,\text{R}}$, given by Eq.~\eqref{eq:T2-timescale}, gives the total decoherence rate of the qubit
\begin{equation}
  \label{eq:T2-timesale-modified}
  \Gamma_2 = \Gamma_{2,\text{R}} + \Gamma_\text{B}\,.
\end{equation}

Assuming the other decoherence factors constant, we can concentrate on the exact nature of the decoherence driven by the resistor. Especially, solving the master equation \eqref{eq:ME!!} using the block diagonalization method allows for the computation of the temperature dependency of the decoherence rate $\Gamma_{2, \text{R}}$ by picking the coherence decay rate from the eigenvalues of the Liouvillian block $\liouv_1$. In practice, the eigenvalues for the Liouvillian block for a specific temperature are computed using a proper Hilbert space truncation size $S$ for that temperature, as explained earlier. Then Eq.~\eqref{eq:T2-timescale} is applied to those eigenvalues. This gives an estimate on how the decay behaves with respect to temperature.

\begin{figure}
    \centering
    \includegraphics[width=\linewidth]{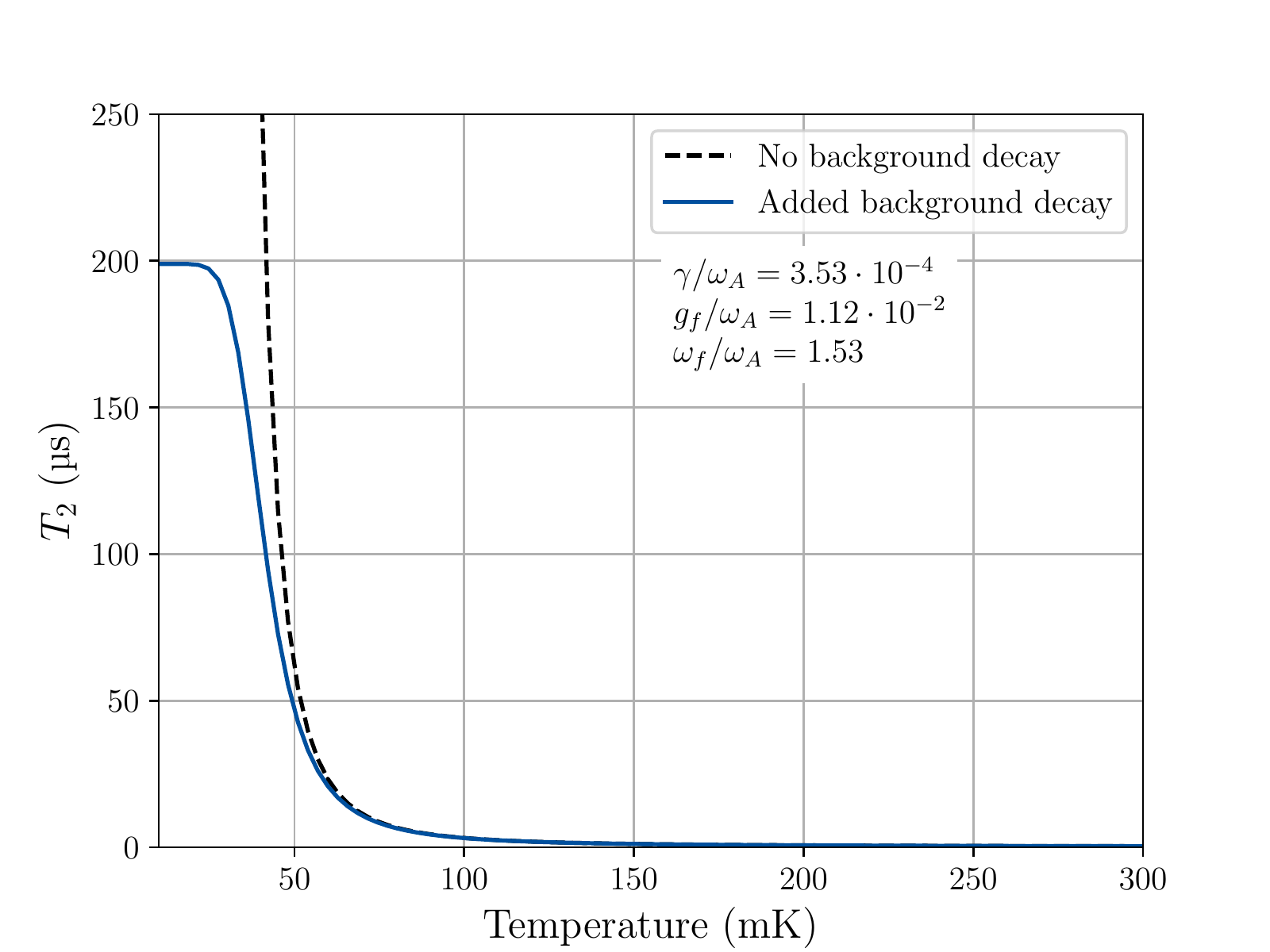}
    \caption{Temperature dependency of the decoherence time with and without an additional background decay rate of $\Gamma_\text{B} = 5026\,\mathrm{s^{-1}}$, which corresponds to a $T_2$ time of about $200\,\mathrm{\mu s}$. The listed parameters have been normalized by the qubit frequency $\omega_\text{A}$.}
    \label{fig:T2-vs-temp1}
\end{figure}

From the decay rates we can obtain the characteristic decoherence times $T_2 = 1/\Gamma_2$. In Fig.~\ref{fig:T2-vs-temp1} the decoherence time scale is plotted as a function of the bath temperature using the same circuit parameters as in Fig.~\ref{fig:real-qubit-coherence-decay}. The decoherence effect coming purely from the resistor is plotted as the black dashed line, which diverges for low temperatures as predicted in Sec.~\ref{sec:decoherenceBlockadeZeroT}. The additional background decay rate from other sources makes the $T_2$ time scale stabilize to a value close to $200\,\mathrm{\mu s}$, as can be seen from the blue line. This behaviour is what is observed experimentally when the temperature of the mixing chamber is reduced \cite{Yeh_2017}. The existence of excess thermal photons that leak to the readout resonator from the attenuators \cite{Yeh_2017, Wang_2019} and other background decay processes can explain this saturation effect.

We observe that the effect of resistor decoherence becomes negligible for low enough temperatures compared to the other sources. The diverging behaviour of the resistor contribution can be explained by noting that near absolute zero the resonator's steady state, the thermal state, approaches the state $\ket{0}\bra{0}$. Thus the absolute value of the real part of the eigenvalues gets smaller, indicating a situation where coherence decays very slowly. At absolute zero the decoherence effect from the resistor vanishes completely because in that case the unitary part of the master equation commutes with the dissipators, leading to purely unitary dynamics when the resonator is in a thermal state, as explained in more detail in Sec.~\ref{sec:decoherenceBlockadeZeroT}.

\subsection{Trade off between $\gamma$ and $g_\text{f}$}

\begin{figure}
    \centering
    \includegraphics[width=\linewidth]{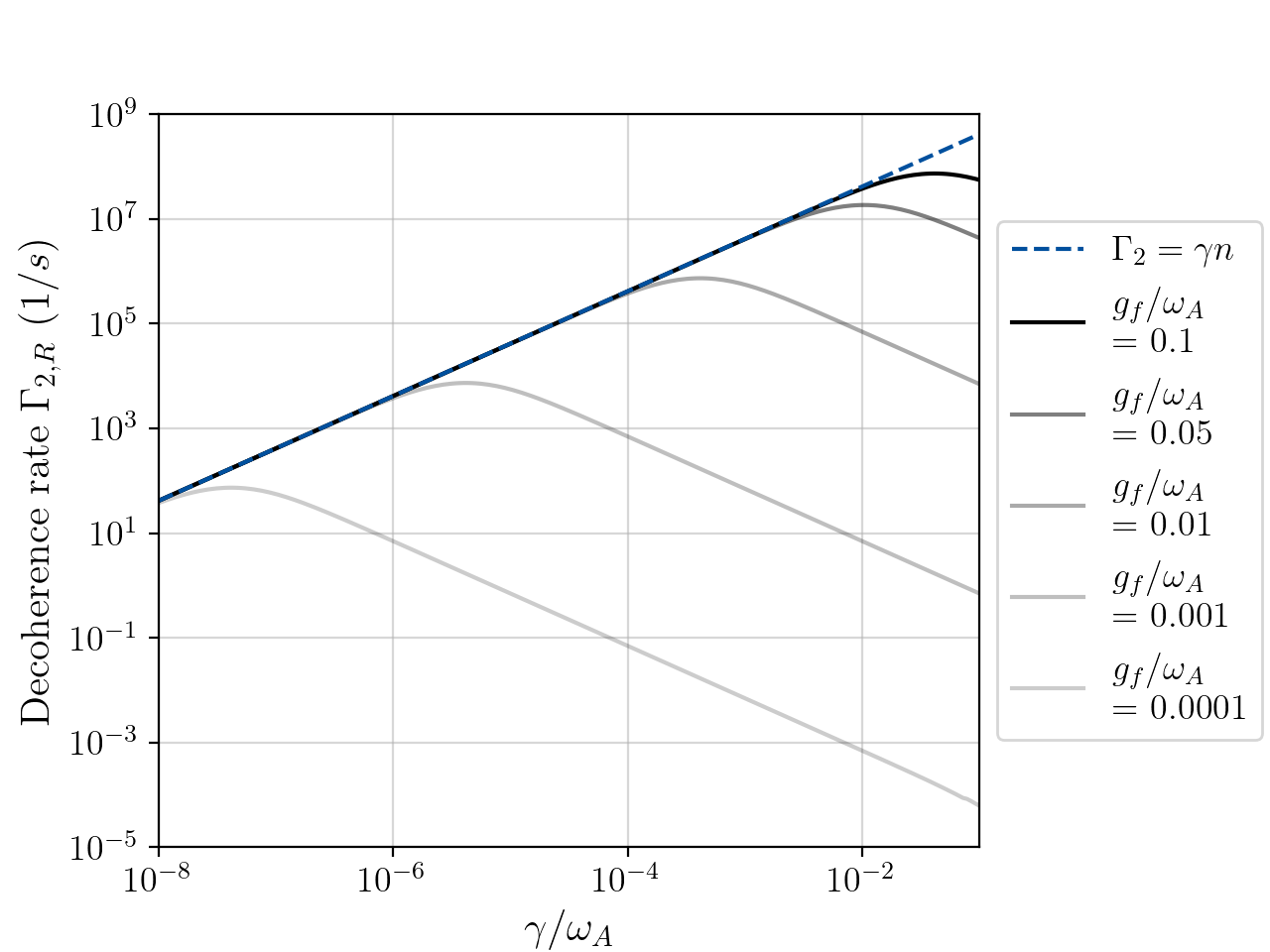}
    \caption{Resistor's contribution $\Gamma_{2, \text{R}}$ to the decoherence rate as a function of resonator dissipation rate $\gamma$ for several qubit--resonator couplings $g_\text{f}$ with resistor temperature $T=150\,\mathrm{mK}$. The blue dashed line shows the often used approximation of decoherence rate being linearly proportional to both $\gamma$ and the average number of thermal photons $\bar{n}$.}
    \label{fig:T2-gamma-gf}
\end{figure}

\begin{figure}
    \centering
    \includegraphics[width=\linewidth]{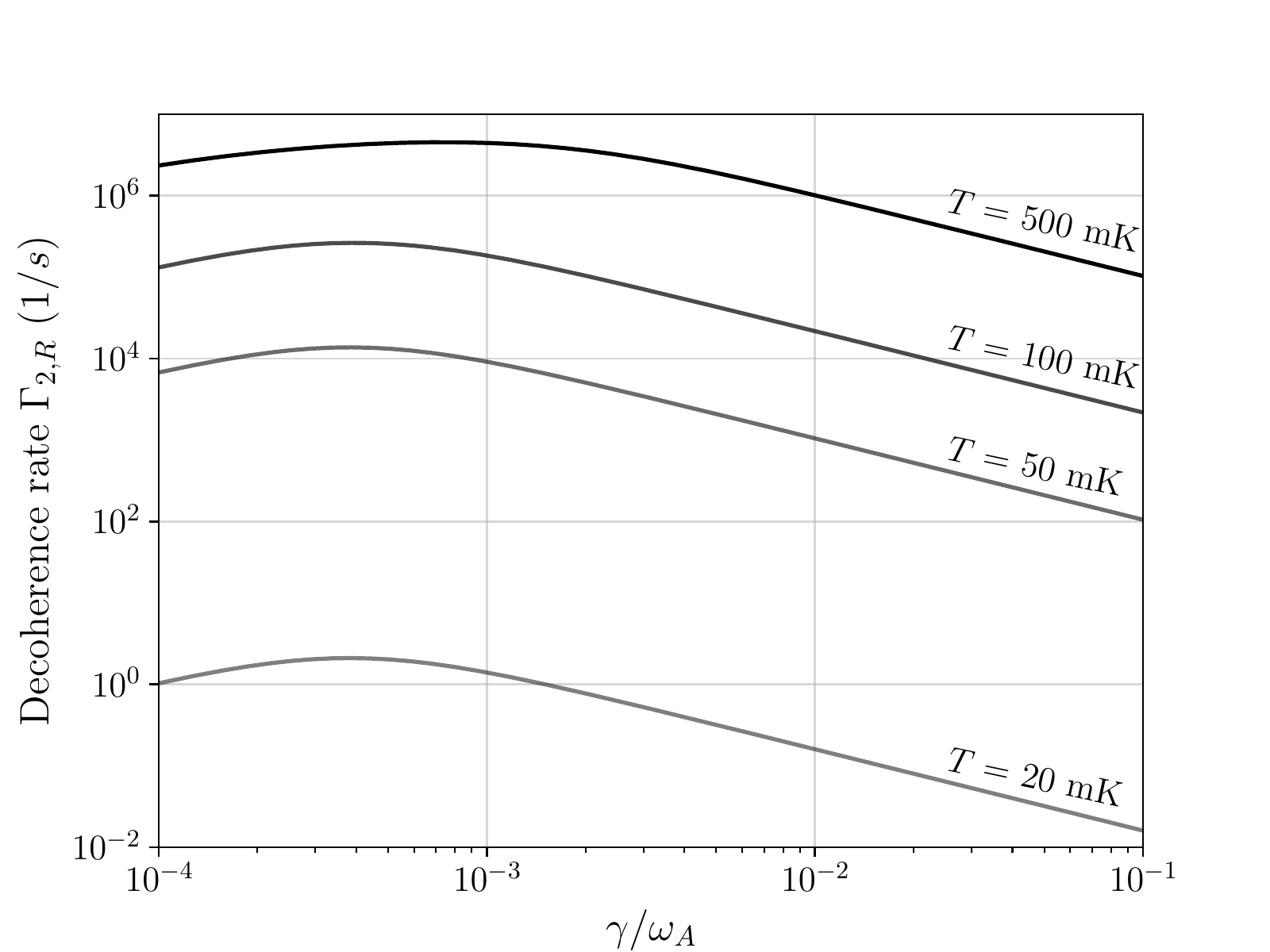}
    \caption{$\Gamma_{2, \text{R}}$ as a function of $\gamma^\prime = \gamma/\omega_\text{A}$ for different temperatures with fixed qubit--resonator coupling of $g_\text{f} = 0.01\omega_\text{A}$.}
    \label{fig:T2-gamma-T}
\end{figure}

Not fixing the coupling parameters to specific values allows us to compute dynamics of the qubit in different coupling regimes. This allows for the study of how changing the coupling parameters affects the time evolution of the qubit and specifically the decoherence rate. This is shown in Figs.~\ref{fig:T2-gamma-gf} and \ref{fig:T2-gamma-T}, where the resistor's contribution to the decoherence rate $\Gamma_{2, \text{R}}$ is plotted versus the resonator dissipation rate $\gamma$ for some values of qubit--resonator coupling $g_\text{f}$ and temperature respectively.

In Fig.~\ref{fig:T2-gamma-gf} we observe how the decoherence rate behaves as a function of $\gamma$, normalized by the qubit frequency (see Eqs.~\eqref{eq:g_f} and \eqref{eq:gamma_prime}), for a specific temperature of $T=150\,\mathrm{mK}$. We can notice that for small values of the ratio $\gamma^\prime/g_\text{f}^{\prime 2}$, the increase in $\gamma^\prime$ increases the decoherence rate of the qubit due to the resistor. In this dispersive strong regime \cite{Schuster_2007} the decoherence rate coincides with the often used approximation of decoherence rate being driven linearly by the number of thermal photons in the resonator \cite{Yeh_2017}, according to
\begin{equation}
  \label{eq:rncm}
  \Gamma_2 = \gamma\bar{n}\,,
\end{equation}
where $\bar{n}$ is the expected number of quanta in the resonator. This approximation is valid when the coupling rates $\gamma$ and $g_\text{f}$ are small enough compared to the resonator's resonance frequency \cite{Yeh_2017} and in addition, when the ratio $\gamma^\prime/g_\text{f}^{\prime 2}$ is small \cite{Clerk_2007} (see Fig.~\ref{fig:T2-gamma-gf}). If the ratio $\gamma^\prime/g_\text{f}^{\prime 2}$ is not small enough, we observe that Eq.~\eqref{eq:rncm} predicts too large coherence decay rates. Remarkably, outside the regime of validity of Eq.~\eqref{eq:rncm} we observe the opposite behavior, that is, increasing the dissipation rate of the resonator actually improves the decoherence rates. One way to explain this is observing that, if we consider the limit $\gamma^\prime/g_\text{f}^{\prime 2}\gg 1$, then the timescale of the resonator-resistor dissipative interaction is much faster than the timescale of the unitary resonator-qubit interaction. This situation may intuitively resemble the well-known quantum Zeno effect \cite{BreuerPetruccione}. Indeed, neglecting momentarily the local unitary dynamics of the resonator and resistor only, we can write the master equation driving the system evolution as:
\begin{equation}
    \label{eqn:MEzeno}
    \mathcal{L}[\rho(t)]=-\i g_\text{f}\lambda[\sigma_z a_\text{f}^\dagger a_\text{f},\rho(t)]+\gamma \mathcal{D}[\rho(t)],
\end{equation}
where the structure of the dissipator $\mathcal{D}$ can be found in Eq.~\eqref{eq:ME!!}. Since $\gamma^\prime/g_\text{f}^{\prime 2}\gg 1$, after a time interval that is almost infinitesimal (in comparison to the timescale given by $g_\text{f}^{\prime-2}$) the system will be driven toward a state $\rho_{QR}$ such that $\mathcal{D}[\rho_{QR}]=0$, i.e., the reduced state of the resonator will be thermal. Then, this master equation looks like the one describing the quantum Zeno effect (e.g., consider Eq.~(3.359) of Ref.~\citenum{BreuerPetruccione}), and the state of the overall system will ``freeze'' in $\rho_{QR}$. Reintroducing the local unitary dynamics of the qubit and resonator, we observe that the qubit dynamics is effectively detached from the one of the resonator. Therefore, the qubit will not feel the presence of a source of dissipation and decoherence (i.e., the resistor), and will oscillate unitarily for any time $t$.

In Fig.~\ref{fig:T2-gamma-T} we can see a similar behaviour as in Fig.~\ref{fig:T2-gamma-gf} but this time the different lines correspond to different temperatures with fixed $g_\text{f}=0.01\omega_\text{A}$. We observe that the resistor's contribution to the decoherence rate is exponentially suppressed if the temperature of the resistor is decreased, as can be seen from the difference of several orders of magnitude in $\Gamma_{2,\text{R}}$ between $T=20\,\mathrm{mK}$ and $T=50\,\mathrm{mK}$. This behavior coincides with what was shown in Fig.~\ref{fig:T2-vs-temp1} for the case of no added background decay rate. We can also notice in Fig.~\ref{fig:T2-gamma-T} that the decoherence rates are lower than what would be predicted by the linear approximation \eqref{eq:rncm}. This approximation is starting to be valid on the left side of the figures, before the maxima of the graphs. It should be noted that changing the value of $g_\text{f}$ changes the position of the maxima as the approximation \eqref{eq:rncm} either becomes better (for lower $\gamma^\prime/g_\text{f}^{\prime 2}$) or worse (for higher $\gamma^\prime/g_\text{f}^{\prime 2}$). Then, we recover the same behavior as in Fig.~\ref{fig:T2-gamma-gf}, where the decoherence rate improves when the resonator dissipation rate $\gamma$ is increased after a certain value, when the ratio $\gamma^\prime/g_\text{f}^{\prime 2}$ is no longer small.

%% file: conclusions.tex
\section{Conclusions \label{sec:concusions}}

We have shown a direct method for obtaining the dynamics of a dispersive transmon qubit in the presence of noise from the qubit control line. Our derivation connects the circuit components of the lumped element model of the circuit directly to the time evolution of the qubit by expressing the coupling parameters of the master equation as functions of the circuit parameters. This allows for exact study of how each of the individual components affects the time evolution of the qubit. We have solved the master equation numerically by setting values to the circuit elements that could correspond to the ones used in a real life implementation of a superconducting circuit. This has allowed us to obtain results that could be directly tested experimentally, by measuring the time evolution of the qubit coherences for different temperatures.

In solving the master equation numerically, we have exploited the symmetries of the Liouvillian superoperator and have been able to transform it into a block diagonal form. This method has been particularly useful because, when studying the decoherence effects, it has allowed us to concentrate only on a small part of the, possibly very large, Liouvillian matrix. This method reduces the number of needed matrix elements from $s^4S^4$ to $s^2S^2$, where $s$ is the truncation of the qubit's Hilbert space (so $s=2$) and $S$ is the size of Hilbert space truncation for the resonator. The block diagonalization of the Liouvillian is also completely general and can be done to any Liouvillian that satisfies the requirements given in \cite{Cattaneo_2020} so that this method can be employed also when discussing different master equations arising from other superconducting circuits.

Starting from the master equation derived in Sec.~\ref{sec:derivation}, we have proven the intuitive result that the contribution to the decoherence rate coming from thermal noise in the drive line vanishes when the resistor temperature goes to zero. Moreover, our numerical analysis (see Fig.~\ref{fig:T2-vs-temp1}) shows that, for values of the circuit parameters that are currently employed in circuit QED laboratories, bringing the resistor temperature below 50mK improves the qubit decoherence rate by several orders of magnitude. This highlights the importance of being able to tune the temperature of the attenuator in the drive line.

In addition, we have explored the qubit decoherence rate for different parameter regimes of $\gamma$ (resonator dissipation rate) and $g_\text{f}$ (qubit--resonator coupling), and the results of our analysis are depicted in Figs.~\ref{fig:T2-gamma-gf} and~\ref{fig:T2-gamma-T}. We see that in the dispersive strong limit, where $\gamma^\prime/g_\text{f}^{\prime 2} \ll 1$ (the primed values are the physical values renormalized by the qubit frequency) our model agrees with the often used approximation of the decoherence rate being linearly proportional to the amount of thermal photons in the resonator. According to this approximation, if the resonator dissipation rate is increased then the decoherence rates get worse. However, we notice a seemingly counter-intuitive behavior when we move away from the dispersive strong regime to dispersive weak regime where $\gamma^\prime/g_\text{f}^{\prime 2} \gtrsim 1$. In this regime we see that the decoherence rates are improving when $\gamma$ is increased. We may explain this phenomenon as follows: if the decay time of the resonator is much shorter than the timescale of the unitary qubit--resonator dynamics, then the state of the overall system ``freezes'', and this is similar to what happens in the quantum Zeno effect, when very fast measurements are performed repetitively on the system. So, in this regime the dynamics of the qubit and of the resonator are effectively detached, and the qubit experiences only its free unitary time evolution. 

Although the weak dispersive limit has the downside that the a.c.~Stark shift is too small to dispersively resolve individual photons, quantum non-demolition experiments of the qubit can still be performed \cite{Schuster_2007}. Moreover, it has recently been shown that high fidelity state preparation in the dispersive weak regime is possible \cite{Eickbusch_2021}. This result combined with the possibility of improved decoherence rates in the dispersive weak regime suggests that more research should be conducted also in this regime of cQED applications in order to push forward the development of superconducting qubits. In particular, our results in Figs.~\ref{fig:T2-gamma-gf} and~\ref{fig:T2-gamma-T} show that the contribution to the qubit decoherence rate from the thermal noise in the drive line may be reduced by several orders of magnitude by working in the dispersive weak regime.

As a next step, the extension of this work to describe drive-line-induced decoherence for transmon qubits coupled to multiple resonator modes \cite{Malekakhlagh-pra-2016,Malekakhlagh-pra-2016b,Malekakhlagh-prl-2017} deserves further investigation. A theoretical treatment beyond the single-mode Rabi model will also be required for description of ultra strong qubit--resonator coupling regimes where higher modes no longer play a weak perturbative role in dynamics \cite{Bosman_2017,gely-prb-2017}. Simulation of systems with engineered frequency-dependent impedances---such as Purcell filtered resonators---is also directly relevant to optimization of fast high-fidelity readout schemes in the presence of noise \cite{walter-prapp-2017}. 

%% file: acknowledgements.tex
\section*{Acknowledgments}
A.V. and M.C.  thank Sabrina Maniscalco for useful discussions about the results of this work.
M.C.~gratefully acknowledges the Spanish State Research Agency through the Severo Ochoa and Mar\'{i}a de Maeztu
Program for Centers and Units of Excellence in
R\&D (MDM-2017-0711) and through the QUARESC
project (PID2019-109094GB-C21 and -C22/ AEI /
10.13039/501100011033). He also acknowledges funding
by CAIB through the QUAREC project (PRD2018/47), and by the Academy of Finland via the Centre of Excellence program (Project No. 336810 and Project No. 336814).

%% file: appendix1.tex
\section{Caldeira-Leggett model for the resistor} \label{appendix2} 

\begin{figure*}
  \center
  \begin{circuitikz}[cute inductors, scale=1]
    \draw (0,0) to[R=$Z$] (0,3);
    \filldraw (0,0) circle [radius=1.5pt];
    \filldraw (0,3) circle [radius=1.5pt];
    \foreach \y in {1.4, 1.5, 1.6}
    \draw (0.6,\y) -- (1.1,\y);
    \filldraw (1.5, 1.5) circle [radius=1.5pt];
    \draw (1.5, 1.5) -- (2, 1.5);
    \draw (2, 1.5) -- (2, 2.5) to[L=$L_1$] (4, 2.5) -- (4, 0.5);
    \draw (2, 1.5) -- (2, 0.5) to[C=$C_1$, v=$\phi_1$] (4, 0.5);
    \draw (4, 1.5) -- (5, 1.5);
    \draw (5, 1.5) -- (5, 2.5) to[L=$L_2$] (7, 2.5) -- (7, 0.5);
    \draw (5, 1.5) -- (5, 0.5) to[C=$C_2$, v=$\phi_2$] (7, 0.5);
    \draw (7, 1.5) -- (7.5, 1.5);
    \draw[dashed] (7.5, 1.5) -- (8.5, 1.5);
    \draw (8.5, 1.5) -- (9, 1.5);
    \draw (9, 1.5) -- (9, 2.5) to[L=$L_n$] (11, 2.5) -- (11, 0.5);
    \draw (9, 1.5) -- (9, 0.5) to[C=$C_n$, v=$\phi_n$] (11, 0.5);
    \draw (11, 1.5) -- (11.5, 1.5);
    \draw[dashed] (11.5, 1.5) -- (12.5, 1.5);
    \draw (12.5, 1.5) -- (13, 1.5);
    \filldraw (13, 1.5) circle [radius=1.5pt];
  \end{circuitikz}  
\caption{The impedance $Z$ of a resistor is represented as an infinite series of LC-oscillators using the extended Foster's first form. Figure adapted from \cite{Cattaneo2021}.}
\label{Fig: LC-resistor}
\end{figure*}
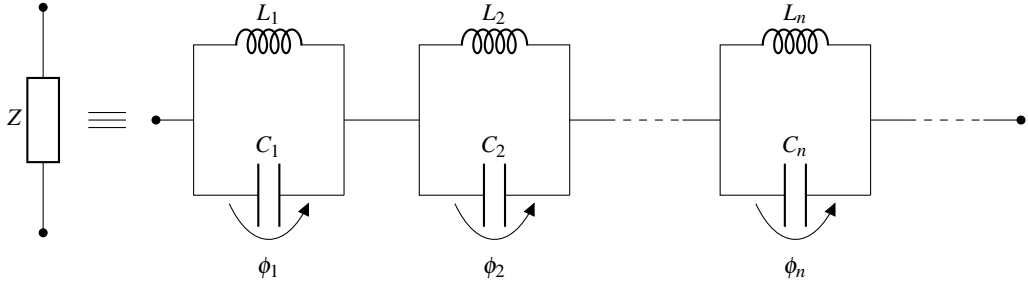

To get a model for quantum dissipation that describes the lossy nature of a resistor, we extend (following the example of \cite{Vool_Devoret2017} and \cite{Cattaneo2021}) a series of parallel LC-oscillators encountered in Foster's first form \cite{Foster} to be infinitely long (see Fig.~\ref{Fig: LC-resistor}). This transforms our model from a simple one degree of freedom model (that is resistor with resistance $R$) into one that has an infinite number of degrees of freedom, which are the node fluxes between each of the parallel LC-oscillators in Fig.~\ref{Fig: LC-resistor}. The change from a finite number of degrees of freedom to an infinite one allows for the emergence of irreversible dynamics of the system on physical timescales \cite{Vool_Devoret2017}.

The first step in the derivation of the model of quantum dissipation is to analyze the impedances in Fig.~\ref{Fig: LC-resistor}. Each of the parallel LC-oscillators have an impedance of 
\begin{equation}
    Z_j(\omega) = \frac{\i}{2C_j}\Bigg(\frac{1}{\omega + \omega_j} + \frac{1}{\omega - \omega_j}\Bigg)\,,
    \label{Eq: Z_LC}
\end{equation}
where $\omega_j$ is the resonance frequency of the $j$th oscillator given by $\omega_j = 1/\sqrt{L_j C_j}$. Although the impedance of the LC-oscillator seems purely complex, the divergence at resonance frequencies $\pm\omega_j$ will also yield a real part for the impedance. To see this, we need to apply the inverse Fourier transform to move to the time domain. However, in order to do this we need to allow the frequency to take complex values as $\omega \rightarrow \omega + \i\epsilon$ for some small $\epsilon \in \mathbb{R}$. In this way the impedance codifies causality and is correctly defined as a boundary distribution in the limit of $\epsilon\to 0^+$. Now the inverse Fourier transform of the impedance function $Z_j(\omega)$ becomes
\begin{equation}
    Z_j(t) = \lim_{\epsilon\to 0^+}\frac{1}{2\pi}\int_{-\infty}^\infty \d\omega \e^{\i\omega t}\e^{-\epsilon t}Z_j(\omega + \i\epsilon)\,.
\end{equation}
Using the expression in Eq.~\eqref{Eq: Z_LC} we get
\begin{align}
    Z_j(t) = \lim_{\epsilon\to 0^+}\frac{\i\e^{-\epsilon t}}{4\pi C_j}\Big[&\int_{-\infty}^\infty \d\omega \frac{\e^{\i\omega t}}{\omega + \i\epsilon + \omega_j} \nonumber \\ 
    &+ \int_{-\infty}^\infty \d\omega \frac{\e^{\i\omega t}}{\omega + \i\epsilon - \omega_j}\Big]\,.
\end{align}
To the above integrals we can use the Sokhotski-Plemelj theorem for the real line
\begin{equation}
    \lim_{\epsilon \to 0^+}\int_a^b\d x\frac{f(x)}{x\pm\i\epsilon} = \mp\i\pi f(0) + \text{P.V.}\int_a^b\d x\frac{f(x)}{x}\,,
    \label{Eq: Sokhotski-Plemelj}
\end{equation}
where $a<0<b$, $f$ is defined and continuous on the integration interval and $\text{P.V.}$ denotes the Cauchy principal value. Applying the theorem yields
\begin{multline}
    Z_j(t) = \frac{1}{4C_j}\left(\e^{\i\omega_j t} + \e^{-\i\omega_j t}\right) \\ + \frac{\i}{4\pi C_j}\text{P.V.}\int_{-\infty}^\infty \d\omega \e^{\i\omega t}\left(\frac{1}{\omega + \omega_j} + \frac{
    1}{\omega - \omega_j}\right)\,.
\end{multline}
We can see that the impedance has now a real term, which becomes a pair of delta functions when we move back to the frequency space via a Fourier transform:
\begin{multline}
    \tilde{Z}_j(\omega) = \frac{\pi}{2}\omega_j z_j [\delta(\omega - \omega_j) + \delta(\omega + \omega_j)] \\ + \frac{\i}{2}\omega_j z_j\left(\frac{1}{\omega + \omega_j} + \frac{1}{\omega - \omega_j}\right)\,,
    \label{Eq: Z_j(omega)-with-real-part}
\end{multline}
where we used the frequency of an oscillator $\omega_j = 1/\sqrt{C_jL_j}$ and the characteristic impedance $z_j = \sqrt{L_j/C_j}$ to write $1/C_j = \omega_j z_j$. 

What Eq.~\eqref{Eq: Z_j(omega)-with-real-part} tells us is that the single LC-oscillator has a point support on the real line, given by the $\delta$-functions. This is indication of non-dissipative nature of the parallel LC-oscillator. When we sum over all of the different LC-oscillators in the Foster's first form, mathematically we build a dense comb of $\delta$-functions, which gives an extended support on the real line and thus is a signature of dissipation.

Setting now $\omega_j = j\Delta\omega$, for $j\in\mathbb{N}$ and $\Delta\omega$ being small frequency step, the impedance of the individual oscillator can be written as
\begin{equation}
    z_j = \frac{2\Delta\omega}{\pi\omega_j}\text{Re}[Z(\omega_j)]\,,
\end{equation}
and the individual capacitance $C_j$ and inductance $L_j$ are then given by
\begin{align}
C_j &= \frac{1}{\omega_j z_j} = \frac{\pi}{2\Delta\omega\text{Re}[Z(\omega_j)]} \,,\\
L_j &= \frac{1}{\omega_j^2C_j} = \frac{2\Delta\omega}{\pi\omega_j^2}\text{Re}[Z(\omega_j)]\,, \label{Eq: L_j}
\end{align}
where $Z(\omega)$ is the impedance of the resistor, that we want to mimic.

We can now write the impedance of the infinite collection of the parallel LC-oscillators as
\begin{equation}
    Z_\infty(\omega) = \sum_{j=0}^\infty\tilde{Z}_j(\omega)\,.
\end{equation}
Taking the limit of $\Delta\omega\to 0$, we can transform the sum into an integral and finally get
\begin{multline}
  Z_\infty(\omega) = \text{Re}[Z(\omega)] + \frac{\i}{\pi}\Bigg[\text{P.V.}\int_0^\infty \d\omega_j\frac{\text{Re}[Z(\omega_j)]}{\omega + \omega_j} \\
  + \text{P.V.}\int_0^\infty \d\omega_j\frac{\text{Re}[Z(\omega_j)]}{\omega - \omega_j}\Bigg]\,.
    \label{Eq: Z_infty PV integrals}
\end{multline}

From the above we can notice that
\begin{align}
  \text{Re}[Z_\infty(\omega)] &= \text{Re}[Z(\omega)] \label{Eq: solved real impedance}\,,\\
  \text{Im}[Z_\infty(\omega)] &= \frac{1}{\pi}\Bigg[\text{P.V.}\int_0^\infty \d\omega_j\frac{\text{Re}[Z(\omega_j)]}{\omega + \omega_j} \nonumber \\
  &+ \text{P.V.}\int_0^\infty \d\omega_j\frac{\text{Re}[Z(\omega_j)]}{\omega - \omega_j}\Bigg]\,.
  \label{Eq: solved imag impedance}
\end{align}

We can now choose the impedance of interest $Z(\omega)$ in Eqs.~\eqref{Eq: solved real impedance} and \eqref{Eq: solved imag impedance} such that it stays approximately constants with respect to frequency, mimicking the behaviour of a real resistor. This behaviour is obtained, for example, by choosing the ohmic spectrum of the resistor \cite{Cattaneo2021}
\begin{equation}
  Z_\text{R}(\omega) = R\frac{\omega_\text{c}^2}{\omega_\text{c}^2 + \omega^2} + \i R\frac{\omega\omega_\text{c}}{\omega_\text{c}^2 + \omega^2}\,,
  \label{eq: Resistor impedance}
\end{equation}
where $R$ is the resistance of the resistor and $\omega_\text{c}$ is some large cutoff frequency such that the above equation is valid in the range $\omega \ll \omega_\text{c}$.

Now, knowing that the resistor can be modelled as a collection of LC-oscillators, we can quantize it by considering the quantization of the separate LC-oscillators that constitute the network. We choose to parametrize the problem as shown in Fig.~\ref{Fig: LC-resistor}, where each of the fluxes $\phi_j$ is associated with the voltage difference over the $j$th LC-circuit \cite{Cattaneo2021}. In this way the LC-circuits are not coupled and can be treated as independent harmonic oscillators with angular frequency $\omega_j$. Applying the quantization procedure of the harmonic oscillator to each LC-oscillator, we get the Hamiltonian of the resistor as
\begin{equation}
    H_\text{R} = \sum_{j=1}^\infty \hbar\omega_j\hat{a}^\dagger_j\hat{a}_j \,.
    \label{Eq: H_r with inf}
\end{equation}
This highlights the fact that the resistor can be treated as a bosonic bath.

%% file: appendix2.tex
\section{Circuit Hamiltonian derivation details} \label{appendix1} 

We begin by considering the capacitive and inductive energies of the circuit in Fig.~\ref{fig:circuit-model} by writing them in terms of the generalized ﬂux variables as usually done in circuit quantum electrodynamics (cQED) \cite{Vool_Devoret2017,Girvin2014}. The generalized flux is voltage through some element $b$ of the circuit integrated with respect to time \cite{Blais_cQED, Vool_Devoret2017}, $\phi_b(t) = \int_{-\infty}^t V_b(t^\prime)\d t^\prime$, where the lower limit of integration denotes a time sufficiently far in the past, such that the circuit was at rest. Usually we can treat the circuit elements in a simple fashion by writing down the energy stored in the element and using that to construct the Lagrangian. However, treating the resistor in this way is not trivial because it is a dissipative circuit element, thus leading to irreversible behavior which is not well treated through unitary quantum mechanics. Fortunately this problem can be overcome by extending the formalism via the usage of the Caldeira-Leggett model \cite{Caldeira-Leggett} in the context of cQED alongside with a network synthesis method called Foster's first form \cite{Foster}.

The full derivation of the Caldeira-Leggett model for the resistor was outlined in appendix \ref{appendix2} and can be found in full detail in reference \cite{Cattaneo2021}. In short summary, we can view the resistor as a bosonic bath consisting of an infinite amount of harmonic oscillators. These harmonic oscillators are essentially parallel LC-oscillators attached in an infinite series, which gives rise to a real impedance that corresponds to the resistance of the resistor. Physically, the signal travelling through the infinite series of parallel LC-oscillators can never reach the end of the chain but is lost travelling down the chain forever \cite{Cattaneo2021}. This is seen as dissipation at the input and allows for the quantum mechanical treatment of the resistor.

The circuit we are interested in contains both capacitive and inductive elements. Most of these elements are linear in nature but the qubit inductive potential $V(\phi_\text{A})$ is left to be an undetermined nonlinear potential, which later gives rise to anharmonicity in the qubit energy levels \cite{Koch2007}. Looking at the circuit diagram in Fig.~\ref{fig:circuit-model}, we can identify the capacitive and inductive energies of the circuit. The capacitive part can be written as
\begin{equation}
  \label{eq:Capacitive-energy}
  E_\text{C} = \frac{1}{2}C_\text{A}\dot{\phi}_\text{A}^2 + \frac{1}{2}C_\text{f}\dot{\phi}_\text{f}^2 + \frac{1}{2}C_\text{g}(\dot{\phi}_\text{A} - \dot{\phi}_\text{f})^2 + \sum_{k=1}^\infty\frac{1}{2}C_k\dot{\phi}_k^2\,,
\end{equation}
while the inductive part can be written as
\begin{multline}
  \label{eq:inductive-energy2}
  E_\text{L} = V(\phi_\text{A}) + \frac{1}{2L_\text{f}} \phi_\text{f}^2 + \frac{M}{L_\text{L}L_\text{f}}\phi_\text{f}\Big(\sum_k^\infty\phi_k\Big) \\ + \frac{1}{2L_\text{L}}\Big(\sum_k^\infty\phi_k\Big)^2 + \sum_{k=1}^\infty\frac{1}{2L_k}\phi_k^2\,.
\end{multline}
In both Eqs.~\eqref{eq:Capacitive-energy} and \eqref{eq:inductive-energy2} the sum over $k$ comes from the resistor contribution. It describes the summation over the different frequency modes of the resistor. See appendix \ref{appendix2} for more details on the resistor quantization.

 To better see the structure of Eqs.~\eqref{eq:Capacitive-energy} and \eqref{eq:inductive-energy2}, we represent the equations in a matrix form. We define a flux vector $\bm{\phi} = [\phi_\text{A}, \phi_\text{f}, \bm{\phi}_k^\top]^\top$, where the vector $\bm{\phi}_k^\top = [\phi_1, \phi_2, ...]$ collects together the resistor fluxes. We can then rewrite the energy Eqs.~\eqref{eq:Capacitive-energy} and \eqref{eq:inductive-energy2} as
\begin{equation}
  \label{eq:Ec-El-in-matrix-format}
  E_\text{C} = \frac{1}{2}\dot{\bm{\phi}}^\top\bm{C}\dot{\bm{\phi}}\,,\hspace{0.5cm} E_\text{L} = \frac{1}{2}\bm{\phi}\bm{L}^{-1}\bm{\phi} + V(\phi_\text{A})\,,
\end{equation}
where the matrices $\bm{C}$ and $\bm{L}^{-1}$ are given by
\begin{equation}
\label{eq:C-matrix}
\bm{C} = \mleft[
\begin{array}{c c | c c c}
  C_\text{A} + C_\text{g} & -C_\text{g} & 0 & 0 & \hdots \\
  -C_\text{g} & C_\text{f} + C_\text{g} & 0 & 0 & \\
  \hline
  0 & 0 & C_1 & 0 & \\
  0 & 0 & 0 & C_2 & \\
  \vdots & & & & \ddots
\end{array}
\mright]\,,
\end{equation}

\begin{equation}
\label{eq:L-matrix}
    \bm{L}^{-1} = \mleft[
\begin{array}{c c | c c l}
  0 & 0 & 0 & 0 & \hspace{-15pt}\hdots \\
  0 & L_\text{f}^{-1} & \frac{M}{L_\text{L}L_\text{f}} & \frac{M}{L_\text{L}L_\text{f}} &\hspace{-15pt} \\
  \hline
  0 & \frac{M}{L_\text{L}L_\text{f}} & L_1^{-1} + L_\text{L}^{-1} & L_\text{L}^{-1} &\hspace{-15pt} \\
  0 & \frac{M}{L_\text{L}L_\text{f}} & L_\text{L}^{-1} & L_2^{-1} + L_\text{L}^{-1} &\hspace{-15pt} \\
  \vdots & & & & \hspace{-15pt}\ddots
\end{array}
\mright]\,.
\end{equation}
In Eqs.~\eqref{eq:C-matrix} and \eqref{eq:L-matrix} we can see a structure that describes the interaction between the circuit parts. The upper left corner of the matrices describes the system comprised of the qubit and the readout resonator while the lower right part is reserved for the resistor. The other two blocks describe the interaction between the system and the resistor.

Next we define a vector $\bm{a} = [0,1]^\top$ and $\bm{e}$ as a vector full of ones. This allows us to write the above matrices in a more concise manner as
\begin{equation}
  \label{eq:simplified-matrices}
  \bm{C} =
  \begin{bmatrix}
    \bm{S}_\text{C} & 0 \\
    0 & \bm{C}_k
  \end{bmatrix}
  \,,\hspace{0.25cm}
  \bm{L}^{-1} =
  \begin{bmatrix}
    \bm{S}_\text{L} & \frac{M}{L_\text{L}L_\text{f}}\bm{a}\bm{e}^\top \\
    \frac{M}{L_\text{L}L_\text{f}}\bm{e}\bm{a}^\top & \bm{L}_k^{-1} + L_\text{L}^{-1}\bm{e}\bm{e}^\top
  \end{bmatrix}\,.
\end{equation}
The matrices $\bm{S}_\text{C}$ and $\bm{S}_\text{L}$ are the qubit-resonator system matrices given by the top left blocks in Eqs.~\eqref{eq:C-matrix} and \eqref{eq:L-matrix} respectively. $\bm{C}_k$ and $\bm{L}^{-1}_k$ are diagonal matrices with the values $C_k$ and $L_k^{-1}$ on their diagonals. 

Before doing the Legendre transformation and obtaining the Hamiltonian describing the circuit, we perform a point transformation of the flux vector $\bm{\phi}$ \cite{Cattaneo2021, Parra_Rodriguez_2018,Paladino2003,Parra-Rodriguez2021}:
\begin{equation}
  \label{eq:Z-transformation}
  \bm{Z} =
  \begin{bmatrix}
    \mathbb{1}_2 & 0 \\
    0 & M_0^{-1/2}\bm{M}_k^{1/2}
  \end{bmatrix}
  \,,\hspace{0.5cm} \bm{M}_k = \bm{L}_k^{-1} + \xi\bm{e}\bm{e}^\top\,.
\end{equation}
Here $M_0$ is a free constant with units of inverse inductance and $\xi$ is a parameter we can choose freely. In the end this transformation gives us a form for the Hamiltonian with no internal interaction for the resistor, which simplifies the calculations later. As we apply this transformation to the flux vector we get a new variable $\bm{z} = \bm{Z}\bm{\phi}$ and the capacitance and inductance matrices transform as $\bm{C}_z = \bm{Z}^{-1}\bm{C}\bm{Z}^{-1}$ and $\bm{L}_z^{-1} =\bm{Z}^{-1}\bm{L}^{-1}\bm{Z}^{-1}$. Doing the transformation and setting $\xi = L_\text{L}^{-1}$ we diagonalize the bottom right part of the inductance matrix, getting rid of internal interactions between the resistor modes. Then we assume that the mode inductances of the resistor $L_k$ (see Eq.~\eqref{Eq: L_j}) are much smaller than the inductance of the coupling $L_\text{L}$, or $L_k \ll L_\text{L}$. This holds in the mathematical limit $\Delta\omega\to 0$, where $L_k \to 0$. Otherwise we can just pick $\Delta\omega$ to be small enough such that the assumption holds. The similar procedure is done in the case of capacitive coupling in Ref.~\citenum{Cattaneo2021}. We then invert the $\bm{M}_k$ matrix from Eq.~\eqref{eq:Z-transformation} by using the Sherman-Morrison formula. Then, assuming the approximation mentioned above, we get $\bm{M}_k^{-1} = \bm{L}_k$ at the zeroth order of the weak coupling limit, which is the order we need to get a first-order resistor-resonator coupling in the circuit Hamiltonian (see the discussion in Appendix~D.1 of Ref.~\citenum{Cattaneo2021} for further details). This in turn yields us the final form for the transformed capacitance and inductance matrices:
\begin{align}
  \bm{C}_z &=
  \begin{bmatrix}
    \bm{S}_\text{C} & 0 \\
    0 & M_0\bm{L}_k\bm{C}_k
  \end{bmatrix}\,, \label{eq:C_z matrix}\\
  \bm{L}^{-1}_z &=
  \begin{bmatrix}
    \bm{S}_\text{L} & \frac{M}{L_\text{L}L_\text{f}}\bm{a}\bm{f}_k^\top \\
    \frac{M}{L_\text{L}L_\text{f}}\bm{f}_k\bm{a}^\top & M_0\mathbb{1}
  \end{bmatrix} \label{eq:L_z matix}\,,
\end{align}
where we have defined a coupling vector $\bm{f}_k = \sqrt{M_0\bm{M}_k^{-1}}\bm{e}$ \footnote{The reader must be aware that divergence issues may arise in the inversion of $\bm{M}_k$, and in the definition of $\bm{f}_k$ and $|\bm{f}_k|^2$, which will play a key role in the expression for the spectral density of the bath. However, Refs.~\citenum{Cattaneo2021,Parra-Rodriguez2021,Parra_Rodriguez_2018} show that these quantities are always well-defined (the discussion in these papers is focused on the capacitive coupling, but all their conclusions apply for the inductive coupling of the circuit we are interested in). Moreover, Ref.~\citenum{Cattaneo2021} shows how the weak coupling limit simplifies the expression for $\bm{M}_k^{-1}$. We refer the reader to the above-mentioned references for further details and for a more rigorous derivation of these results.}, which describes the coupling of resistor modes to the system.  
Under the transformation the Lagrangian becomes
\begin{equation}
  \label{eq:transformed-lagrangian}
  \mathcal{L} = \frac{1}{2}\dot{\bm{z}}^\top\bm{C}_z\dot{\bm{z}} - \frac{1}{2}\bm{z}^\top\bm{L}_z^{-1}\bm{z} - V(z_\text{A})\,.
\end{equation}
To this Lagrangian we can now perform the Legendre transformation and get the Hamiltonian as
\begin{equation}
  \label{eq:circuit-hamiltonian-matrix}
  H = \frac{1}{2}\dot{\bm{p}}^\top\bm{C}_z^{-1}\dot{\bm{p}} + \frac{1}{2}\bm{z}^\top\bm{L}_z^{-1}\bm{z} + V(z_\text{A})\,.
\end{equation}
Inverting $\bm{C}_z$ from Eq.~\eqref{eq:C_z matrix} and opening up the terms explicitly gives us the  classical Hamiltonian of the circuit in Eq.~\eqref{eq:circuit-Hamiltonian}.

\subsection{Quantization procedure}

The free terms of the Hamiltonian in Eq.~\eqref{eq:circuit-Hamiltonian} correspond to harmonic oscillators as they are of the form $p^2 + z^2$ for coordinate $z$ and conjugate momentum $p$. In the quantization procedure we can therefore promote the variables to corresponding operators, apply the canonical quantization relation $[z,p] = \i\hbar$ and identify the creation and annihilation operators. For the readout resonator part this gives the newly promoted operators in terms of the ladder operators as
\begin{equation}      
    \label{eq:zf-with-ladder-ops}
    z_\text{f} = \sqrt{\frac{\hbar(C_\text{A} + C_\text{g})}{2D\omega_\text{f}}}(a^\dagger_\text{f} + a_\text{f})\,,
\end{equation}
\begin{equation}
    p_\text{f} = \i\sqrt{\frac{\hbar D\omega_\text{f}}{2(C_\text{A} + C_\text{g})}}(a^\dagger_\text{f} - a_\text{f})\,.
    \label{eq:pf-with-ladder-ops}
\end{equation}
The resistor can be treated in an equivalent manner. The operators for the resistor become
\begin{equation}
    \label{eq:z-alpha-with-ladders}
    z_k = \sqrt{\frac{\hbar\omega_k}{2M_0}}(a^\dagger_k + a_k)\,,
\end{equation}
\begin{equation}
\label{eq:p-alpha-with-ladder}
     p_k = \i\sqrt{\frac{\hbar M_0}{2\omega_k}}(a^\dagger_k - a_k)\,.
\end{equation}
For the qubit we expand the cosine transmon potential \cite{Koch2007} as a Taylor series and truncate it in the second order of $z_\text{A}$. This gives the qubit Hamiltonian as
\begin{equation}
  \label{eq:qubit-Ham}
  H_\text{Qubit} = \frac{p_\text{A}^2}{2}\frac{C_\text{f} + C_\text{g}}{D} + \frac{E_\text{J}}{2}\Big(\frac{2\pi}{\phi_0}\Big)^2z_\text{A}^2\,.
\end{equation}
When this is quantized, the ladder operators for the qubit become
\begin{equation}
  \label{eq:za-with-ladders}
  z_\text{A} = \sqrt{\frac{\hbar(C_\text{f} + C_\text{g})}{2D\omega_\text{A}}}(a_\text{A}^\dagger + a_\text{A})\,,
\end{equation}
\begin{equation}
\label{eq:pa-with-ladders}
    p_\text{A} = \i\sqrt{\frac{\hbar D\omega_\text{A}}{2(C_\text{f} + C_\text{g})}}(a_\text{A}^\dagger - a_\text{A})\,.
\end{equation}
Using the results from Eqs.~\eqref{eq:zf-with-ladder-ops} to \eqref{eq:pa-with-ladders}, we can write the quantized form of the circuit Hamiltonian \eqref{eq:circuit-Hamiltonian} in terms of the ladder operators as
\begin{equation}
  \label{eq:quantizedHamiltonian}
  \begin{aligned}
    H &= \underbrace{\frac{1}{2}\hbar\omega_\text{A}\sigma_z}_{\text{Qubit}} + \underbrace{\hbar\omega_\text{f}a^\dagger_\text{f}a_\text{f}}_{\text{Resonator}} + \underbrace{\sum_{k=1}^\infty\hbar\omega_k a_k^\dagger a_k}_{\text{Resistor}} \\
    &- \underbrace{\frac{\i}{2}\hbar C_\text{g}\sqrt{\frac{\omega_\text{A}\omega_\text{f}}{(C_\text{f} + C_\text{g})(C_\text{A} + C_\text{g})}}\sigma_y(a^\dagger_\text{f} - a_\text{f})}_{\text{Qubit-resonator interaction}} \\
    &+ \underbrace{\frac{M}{L_\text{L}}\sum_{k=1}^\infty\sqrt{\frac{\hbar\omega_k}{2}}\sqrt{\frac{\hbar(C_\text{A} + C_\text{g})}{2D\omega_\text{f}}}\frac{\sqrt{L_k}}{L_\text{f}} (a^\dagger_k + a_k)(a^\dagger_\text{f} + a_\text{f})}_{\text{Resistor-resonator interaction}}\,.
  \end{aligned}
\end{equation}
Here we used the definition of the coupling vector $\bm{f}_k = \sqrt{M_0\bm{M}_k^{-1}}\bm{e}$ to write the element of this vector as $f_k = \sqrt{M_0L_k}$. The ladder operators of the qubit Hilbert space were also truncated to two dimensions, giving the Pauli matrices. 

Next we need to use expressions for the relevant quantities of the Caldeira-Leggett model for the resistor, which are introduced in Appendix~\ref{appendix2}. The inductances $L_k$ are given by Eq.~\eqref{Eq: L_j}. The resistor spectrum is given by \eqref{eq: Resistor impedance}. Using these equation we can write the resistor-resonator interaction term of the quantized Hamiltonian as
\begin{equation}
  \label{eq:Hint-term}
  H_\text{int} = \frac{M}{L_L}\mu\sum_{k=1}^\infty\sqrt{\frac{\hbar\Delta\omega R\omega_\text{c}^2}{\pi\omega_k(\omega_\text{c}^2 + \omega_k^2)L_\text{f}^2}}(a_k^\dagger + a_k)(a_\text{f}^\dagger + a_\text{f})\,,
\end{equation}
where introduced a factor $\mu$ as
\begin{equation}
  \label{eq:mu}
  \mu \equiv \sqrt{\frac{\hbar(C_\text{A} + C_\text{g})}{2D\omega_\text{f}}}\,.
\end{equation}
The square root term in the sum in Eq.~\eqref{eq:Hint-term} describes the current fluctuations in the circuit for different frequencies $\omega_k$. This allows us to define a coupling coefficient $h_k$ that describes the interaction strength between the mode $k$ of the resistor and the readout resonator as
\begin{equation}
  \label{eq:g_k}
  h_k = \sqrt{\frac{\hbar\Delta\omega R\omega_\text{c}^2}{\pi\omega_k(\omega_\text{c}^2 + \omega_k^2)}}\frac{\mu}{L_\text{f}}\,.
\end{equation}
The factor $\Delta\omega$ is the (physically fictitious) difference between the frequencies of the Caldeira-Leggett model, and in the limit of $\Delta\omega\rightarrow 0^+$ it becomes a differential that can be used to compute the spectral density in Eq.~\eqref{eq:spectral-density}, which is a well-defined physical quantity. Using these definitions allows us to write the quantized Hamiltonian \eqref{eq:quantizedHamiltonian} in its final form in Eq.~\eqref{eq:Quantized-Circuit-Hamiltonian-final}.

%% file: appendix3.tex
\section{Diagonalization of the system Hamiltonian} \label{appendix3}

We start the diagonalization by introducing the qubit jump operators $\sigma_+ = \ket{e}\bra{g}$ and $\sigma_- = \ket{g}\bra{e}$. We can express the Pauli operators in terms of these two operators as
\begin{align}
  \sigma_z &= \ket{e}\bra{e} - \ket{g}\bra{g} = [\sigma_+,\ \sigma_-]\,, \label{eq:sigma_z}\\
  \sigma_x &= \ket{e}\bra{g}\ + \ket{g}\bra{e}\ = \sigma_+ + \sigma_-\,, \label{eq:sigma_x}\\
  \sigma_y &= \i(\ket{g}\bra{e} - \ket{e}\bra{g}) = \i(\sigma_- - \sigma_+)\,. \label{eq:sigma_y}
\end{align}
Using \eqref{eq:sigma_y} in the system Hamiltonian $H_\text{Sys} = 1/2\hbar\omega_\text{A}\sigma_z + \hbar\omega_\text{f}a^\dagger_\text{f}a_\text{f} - \i\hbar g_\text{f}\sigma_y(a^\dagger_\text{f} - a_\text{f})$ gives us the following Hamiltonian
\begin{equation}
  \label{eq:JCM-before-RWA}
  H_\text{System} = \frac{1}{2}\hbar\omega_\text{A}\sigma_z + \hbar\omega_\text{f}a^\dagger_\text{f}a_\text{f} + \hbar g_\text{f}(\sigma_- - \sigma_+)(a^\dagger_\text{f} - a_\text{f})\,.
\end{equation}
If we now open up the interaction part we get terms proportional to $\sigma_-a^\dagger_\text{f}$, $\sigma_-a_\text{f}$, $\sigma_+a^\dagger_\text{f}$ and $\sigma_+a_\text{f}$. In the interaction picture these terms would oscillate at different frequencies with the following time dependency \cite{gerry_knight_2004}
\begin{equation}
  \label{eq:time-dependencie-of-terms}
\begin{aligned}
  \sigma_-a^\dagger_\text{f} \sim & \ \e^{-\i(\omega_\text{A} - \omega_\text{f})t} \\
  \sigma_-a_\text{f} \sim & \ \e^{-\i(\omega_\text{A} + \omega_\text{f})t} \\
  \sigma_+a^\dagger_\text{f} \sim & \ \e^{\i(\omega_\text{A} + \omega_\text{f})t} \\
  \sigma_+a_\text{f} \sim & \ \e^{\i(\omega_\text{A} - \omega_\text{f})t} \,.
\end{aligned}
\end{equation}
In the above equation two terms, second and third, oscillate faster compared to the other two, first and last. The standard way of justifying of dropping the fast rotating terms is to assume that $\omega_\text{A} + \omega_\text{f} \gg |\omega_\text{A} - \omega_\text{f}|$, which requires that the qubit and resonator frequencies are sufficiently close to each other \cite{Zueco_2009}. This is called the rotating wave approximation. In our case we cannot apply the approximation $\omega_\text{A} + \omega_\text{f} \gg |\omega_\text{A} - \omega_\text{f}|$ as we are going to be working in the dispersive regime. However, it can be shown  that the Hamiltonian (C4) can be represented as a continued fraction with respect to the coupling parameter $g_\text{f}$ \cite{Tur_2000}. The equations equal to the rotating wave approximation are obtained as a correction of first order in $g_\text{f}^2$ and valid for $g_\text{f}/\omega_\text{A} \ll 1$ \cite{Tur_2000,He_2012}, yielding
\begin{equation}
  \label{eq:JCM-Hamiltonian}
  H_\text{System} = \frac{1}{2}\hbar\omega_\text{A}\sigma_z + \hbar\omega_\text{f}a^\dagger_\text{f}a_\text{f} + \hbar g_\text{f}(\sigma_-a^\dagger_\text{f} + \sigma_+a_\text{f})\,.
\end{equation}
This allows us to consider regimes where $\omega_\text{A} + \omega_\text{f} \gg |\omega_\text{A} - \omega_\text{f}|$ does not necessarily hold, such as the dispersive regime, which is needed in order to get this Hamiltonian into a diagonal form. There we assume that the coupling $g_\text{f}$ between the qubit and the resonator is weak compared to the detuning $\Delta = |\omega_\text{A} - \omega_\text{f}|$ of the resonance frequencies between the same elements. To move to the dispersive regime, we introduce a dispersive parameter $\lambda = g_\text{f}/\Delta$ and require that $\lambda \ll 1$. Following Ref.~\cite{Zueco_2009} we apply a unitary transformation $U$ to the system Hamiltonian \eqref{eq:JCM-Hamiltonian}, where the transformation is parametrized by $\lambda$:
\begin{equation}
  \label{eq:unitary-trafo-for-JCM}
  U = \e^{\lambda(\sigma_+a_\text{f} - \sigma_-a^\dagger_\text{f})}\,.
\end{equation}
After applying the unitary transformation we arrive at the dispersive Hamiltonian for the system
\begin{equation}
  \label{eq:dispersive-hamiltonian}
  \begin{aligned}
  H_\text{Sys}^\text{D} &= \frac{1}{2}\hbar(\omega_\text{A} + 2 g_\text{f}\lambda)\sigma_z + \hbar(\omega_\text{f} + g_\text{f}\lambda\sigma_z)a_\text{f}^\dagger a_\text{f} \\ 
  &+ \hbar g_\text{f}\lambda\sigma_-\sigma_+ + \mathcal{O}(\lambda^2)\,,
  \end{aligned}
\end{equation}
where the superscript $\text{D}$ denotes dispersive frame.

We can now neglect terms of the order $\mathcal{O}(\lambda^2)$ and recover a Hamiltonian whose eigenstates are qubit-resonator product states. Therefore the Hamiltonian is diagonal in the basis $\{\ket{e,\ n-1}, \ket{g,\ n}\}$ with $n=0,1,\ldots$, which was chosen to get a more readable form for the eigenenergies in Eq.~\eqref{eq:eigenenergies}. Next we apply the dispersive transformation \eqref{eq:unitary-trafo-for-JCM} to the other terms of the Hamiltonian \eqref{eq:Quantized-Circuit-Hamiltonian-final} and get the full dispersive description of the dynamics given by Eqs.~\eqref{eq:Disperisve-Hamiltonian-in-pieces} to \eqref{eq:H-D-interaction}.

%% file: appendix4.tex
\section{Derivation of the master equation} \label{appendix4}

In the standard microscopic derivation of the Lindblad master equation (see for example \cite{BreuerPetruccione}) we decompose the interaction Hamiltonian between the system and environment into a product form
\begin{equation}
  H_\text{Int} = \alpha\sum_\beta A_\beta\otimes B_\beta\,,
  \label{eq:H_I decomposition}
\end{equation}
where the operators $A_\beta$ act on the Hilbert space of the system of interest and the operators $B_\beta$ act on the Hilbert space of the environment. In our case we have $A_\beta = a^\dagger_\text{f} + a_\text{f} + \lambda\sigma_x$ and $B_\beta = \sum_{k=1}^\infty h_k (a^\dagger_k + a_k)$ (see Eq.~\eqref{eq:H-D-interaction}).The coefficient $\alpha$ describes the interaction strength and is in our case $\alpha = M/L_\text{L}$. Now, due to the diagonalization of the system Hamiltonian as done in Eq.~\eqref{eq:H-D-system}, we can present it in a form
\begin{equation}
  H_\text{Sys}^\text{D} = \sum_i \epsilon_i\ket{\epsilon_i}\bra{\epsilon_i}\,,
  \label{eq:H_S diagonalization}
\end{equation}
where the energy eigenstates $\ket{\epsilon_i}$ are the product states from the set $\{\ket{e,\ n-1}, \ket{g,\ n}\}$ and the corresponding eigenenergies $\epsilon_i$ are given by Eq.~\eqref{eq:eigenenergies}. The diagonalization of the system Hamiltonian is needed in order to construct the jump operators from the system operators $A_\beta$ \cite{BreuerPetruccione}. In general, we can write the system operators as
\begin{equation}
  A_\beta = \sum_{i, j}\ket{\epsilon_i}\braket{\epsilon_i|A_\beta|\epsilon_j}\bra{\epsilon_j} = \sum_{i, j}\braket{\epsilon_i|A_\beta|\epsilon_j}\ket{\epsilon_i}\bra{\epsilon_j}\,.
  \label{eq:A_beta decomposition}
\end{equation}
Fixing the energy difference between the two basis states to be $\epsilon_j - \epsilon_i = \hbar\omega$, with $\omega$ being the Bohr frequency, gives
\begin{equation}
  A_\beta(\omega) = \sum_{\epsilon_j - \epsilon_i = \hbar\omega}\braket{\epsilon_i|A_\beta|\epsilon_j}\ket{\epsilon_i}\bra{\epsilon_j}\,, \hspace{0.15cm} A_\beta = \sum_\omega A_\beta(\omega)\,,
  \label{eq:A_beta(omega)}
\end{equation}
where $A_\beta(\omega)$ are called the jump operators, which change the energy of the state they operate to by $\hbar\omega$.

Let's now compute the jump operators in our case. We have to consider three distinct cases to get a general description for the different jump operators. These three cases come from the three distinct frequency jumps that can appear in our qubit-resonator system. One jump operator will correspond to the qubit staying in the ground state while the amount of quanta in the resonator varies. In the second case the qubit stays excited while the resonator loses or gains quanta. The third case considers the number of quanta in the resonator staying constant while the qubit switches between the excited and ground states. The jump operators arising from the consideration of these three cases can be written as
\begin{align}
  A(E_{g,n} - E_{g,n+1}) &= \sum_{n=0}^\infty\sqrt{n+1}\ket{g,n+1}\bra{g,n}\,, \label{eq:yks}\\
  A(E_{g,n+1} - E_{g,n}) &= \sum_{n=0}^\infty\sqrt{n+1}\ket{g,n}\bra{g,n+1}\,, \label{eq:kaks}\\
  A(E_{e,n} - E_{e,n+1}) &= \sum_{n=0}^\infty\sqrt{n+1}\ket{e,n+1}\bra{e,n}\,, \label{eq:kol}\\
  A(E_{e,n+1} - E_{e,n}) &= \sum_{n=0}^\infty\sqrt{n+1}\ket{e,n}\bra{e,n+1}\,, \label{eq:nel}\\
  A(E_{g,n} - E_{e,n}) &= \lambda\sum_{n=0}^\infty\ket{e,n}\bra{g,n}\,, \label{eq:vii}\\
  A(E_{e,n} - E_{g,n}) &= \lambda\sum_{n=0}^\infty\ket{g,n}\bra{e,n}\,. \label{eq:kuu}
\end{align}
Above we have six jump operators because each of the three cases has a positive and negative frequency associated with them. It can be seen from the above equations that the negative frequency operators correspond to the adjoint of the positive frequency jump operators, $A(-\omega) = A^\dagger(\omega)$. For simplicity we denote the energy differences in Eqs.~\eqref{eq:yks} to \eqref{eq:kuu} as $E_{g,n} - E_{g,n+1} \equiv \hbar\omega_{gg}$, $E_{e,n} - E_{e,n+1} \equiv \hbar\omega_{ee}$, $E_{g,n} - E_{e,n} \equiv \hbar\omega_{ge}$ and $E_{e,n} - E_{g,n} \equiv \hbar\omega_{eg} = -\hbar\omega_{ge}$.

Now we need to consider the validity of the secular approximation in the derivation of the master equation. In the standard microscopical derivation of the Lindblad master equation, the secular approximation deals with the rotating $\e^{\i(\omega-\omega^\prime)t}$ factors of the master equation, which arise from the different frequencies of the jump operators $\omega$ and $\omega^\prime$. In general, we can neglect the contribution of these rotating terms if the frequency difference between them is large enough such that \cite{Cattaneo_2019}
\begin{equation}
  \exists\, t^*\,\text{such that}\,|\omega - \omega^\prime|^{-1} \ll t^* \ll \tau_\text{S} = \mathcal{O}(\hbar^2/\alpha^2)\,,
  \label{eq:coarse graining 2}
\end{equation}
where $t^*$ is some timescale and $\alpha$ describes the interaction strength. The timescales we are working with can be computed by using the eigenenergy Eq.~\eqref{eq:eigenenergies} and they turn out to be
\begin{align}
  \label{eq:freq-diffs}
  |\omega_{gg} - \omega_{ee}| &= 2g_\text{f}\lambda\,, \\
  |\omega_{gg} - \omega_{eg}| &= \Delta + 2(n+1)g_\text{f}\lambda\,, \\
  |\omega_{ee} - \omega_{eg}| &= \Delta + 2ng_\text{f}\lambda\,.
\end{align}
Here $\Delta = |\omega_\text{A} - \omega_\text{f}|$ is the detuning between the qubit and the readout resonator resonance frequencies. Since we are working in the dispersive limit where $\Delta$ is taken to be sufficiently large, we can safely say that crossterms of the master equation dissipator with $A(\omega_{gg/ee})\rho A(\omega_{eg})$ are negligible and can be disregarded. However, we need to be more careful with the other crossterms of the forms $A(\omega_{gg})\rho A(\omega_{ee})$. These cannot be disregarded as the timescale determined by Eq.~\eqref{eq:freq-diffs} is quite large since $\lambda$ is small, which could violate the condition in Eq.~\eqref{eq:coarse graining 2}.

To circumvent this problem, we set $\lambda=0$ (i.e., no direct coupling between qubit and resonator) during the derivation of the master equation, while keeping the direct interaction with $\lambda\neq 0$ in the unitary part of the dynamics driven by the system Hamiltonian. This is a well-known procedure to get a so-called local master equation, which will provide us a valid approximation of the system dynamics if the coupling between the qubit and resonator is sufficiently small \cite{Trushechkin2016,Hofer2017,Scali2020,Cattaneo_2019}. Under this framework, the frequencies associated with the jump operators are $\omega_{gg} = \omega_{ee} = \omega_\text{f}$, so we get rid of all the rotating terms in the master equation derivation according to Eq.~\eqref{eq:freq-diffs}. Thus we are able to obtain a master equation which complies with the secular approximation. Now the master equation can be written as a sum of the dissipators, each arising from a different combination of the jump operators, which after some straightforward algebra gives us
\begin{equation}
  \label{eq:ME-almosttehre}
  \begin{aligned}
    \dot{\rho}_\text{Sys} = &-\frac{\i}{\hbar}\big[H_\text{Sys} + H_{\text{LS}}, \rho_\text{Sys}\big] \\
    &+ \frac{\alpha^2}{\hbar^2}\gamma(\omega_\text{f})\Big[a_\text{f}{\rho}_\text{Sys}a^\dagger_\text{f} - \frac{1}{2}\big\{a^\dagger_\text{f}a_\text{f}, {\rho}_\text{Sys}\big\}\Big] \\
    &+ \frac{\alpha^2}{\hbar^2}\gamma(-\omega_\text{f})\Big[a^\dagger_\text{f}{\rho}_\text{Sys}a_\text{f} - \frac{1}{2}\big\{a_\text{f}a^\dagger_\text{f}, {\rho}_\text{Sys}\big\}\Big]\,.
  \end{aligned}
\end{equation}
Last step left to do is to compute the coefficients $\gamma(\pm\omega_\text{f})$, which are defined as the Fourier transform of the bath correlation function \cite{BreuerPetruccione}
\begin{equation}
    \gamma(\omega) = \int_{-\infty}^\infty\d\tau\langle\tilde{B}^\dagger(\tau)\tilde{B}(0)\rangle\e^{\i\omega\tau}\,.
\end{equation}
This computation yields the final form of the master equation \eqref{eq:ME!!}. The coefficient $\gamma$ describes the coupling strength between the qubit-resonator system and the resistor as an environment. It is dependent on the spectral density of the bath as $\gamma = 2\pi\alpha^2J(\omega_\text{f})$. The expected number of quanta in the resonator $\bar{n}$ follows the Bose-Einstein distribution. The Lamb shift Hamiltonian $H_\text{LS}$ renormalizes only the resonator frequency so it does not affect the dynamics of the qubit in a significant way.

%% file: appendix5.tex
\section{Block structure of the Liouvillian} \label{appendix5}

In the Liouville space the Liouvillian superoperator is written in the form of Eq.~\eqref{eq:Liouvillian-matrixified}. Motivated by the work done in Ref.~\cite{Cattaneo_2020}, we will examine the observable associated with the total number of particles in a system. In our case we need to compute separately the qubit quanta and the resonator quanta. Then the total number of particles operator $N$ is
\begin{equation}
  \label{eq:total-number-of-particles-op}
  N = \sigma_z +  a_\text{f}^\dagger a_\text{f}\,.
\end{equation}
From the above we can define the superoperator $\mathcal{N} = [N, \cdot]$, which can be written as a supercommutator
\begin{equation}
  \label{eq:total-number-of-particles-superop}
  \mathcal{N} = N\otimes\mathbb{1} - \mathbb{1}\otimes N^\top \,.
\end{equation}
Ref.~\cite{Cattaneo_2020} shows that $[\mathcal{N},\mathcal{L}]=0$ for a broad family of master equations if the secular approximation is properly applied. It is easy to see that these two superoperators do commute also for the master equation~\eqref{eq:ME!!}, therefore we can block diagonalize the Liouvillian matrix using the eigenbasis of $\mathcal{N}$.

\begin{figure}
\resizebox{0.47\textwidth}{!}{%
  \begin{tikzpicture}
  
    \matrix[matrix of math nodes,
      left delimiter=(,
      right delimiter=),
      column sep=-0.5pt,
      row sep=-0.5pt,
      every node/.style={minimum width=1.5cm, minimum height=0.75cm},
      G/.style={draw=black, line width=0.5pt, fill=black!20, rounded corners},
      R/.style={draw=myred, line width=0.5pt, fill=myred!20, rounded corners},
      B/.style={draw=myblue, line width=0.5pt, fill=myblue!20, rounded corners},
      ampersand replacement = \&]
      (L){
      |[G]| $\mathcal{L}_{0}$ \&  \&  \&  \&  \& \\
       \& |[B]| $\mathcal{L}_{1}$ \&  \&  \&  \&  \\
       \&  \& |[R]| $\mathcal{L}_{-1}$ \&  \&  \&  \\
       \&  \&  \& |[B]| $\mathcal{L}_2$ \&  \&  \\
       \&  \&  \&  \& |[R]| $\mathcal{L}_{-2}$ \&  \\
       \&  \&  \&  \&  \&  $\ddots$ \\};

     \node[yshift=10pt] (a) at (L-1-1.north) {\footnotesize$\ket{n}\otimes\ket{m}$};
     \node[yshift=1pt] at (a.north) {\scriptsize$d=n-m=0$};

     \node[yshift=7pt] at (L-2-2.north) {\scriptsize$d=1$};
     \node[yshift=7pt] at (L-3-3.north) {\scriptsize$d=-1$};
     \node[yshift=7pt] at (L-4-4.north) {\scriptsize$d=2$};
     \node[yshift=7pt] at (L-5-5.north) {\scriptsize$d=-2$};

     \node[xshift=20pt, yshift=-10pt, above right] at (L-1-1.north east) {\footnotesize Example for 2D truncation of the resonator:};
     \node[above right] (b) at (L-2-2.north east) {\footnotesize Basis states in block $\mathcal{L}_1$ are $\ket{g1}\otimes\ket{g0},$};
     \node[yshift=5pt, xshift=20pt, below right] at (b.south west) {\footnotesize $\ket{e0}\otimes\ket{g0}, \ \ket{e1}\otimes\ket{g1}$, $\ket{e1}\otimes\ket{e0}$};
    
   \end{tikzpicture}
}
\caption{Block structure of the Liouvillian superoperator in the Liouville space. The exact order of the blocks can vary, what is important is that each block is labeled by $d$ and contains only basis elements where the difference in quanta between the basis element kets is $d$. In the given example the resonator is truncated to 2D giving access only to states $\ket{0}$ and $\ket{1}$. The qubit ground $\ket{g}$ corresponds to $0$ and excited state $\ket{e}$ to $1$. In block $\mathcal{L}_1$ all the left side kets have one excitation more than the right side kets.}
\label{fig:block-form-of-L}
\end{figure}
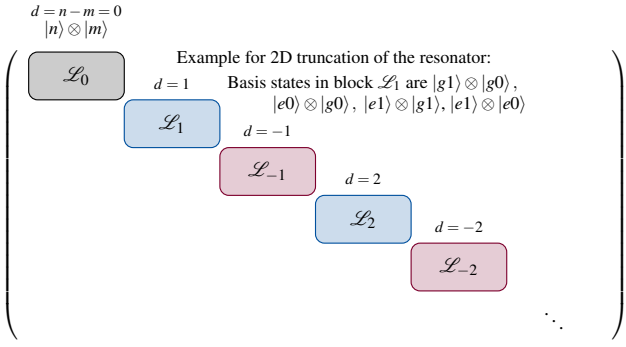

In this block form the different blocks are labelled by the eigenvalues $d$ of the number superoperator $\mathcal{N}$ as expressed in Fig.~\ref{fig:block-form-of-L}. These eigenvalues are given by the difference in the number of quanta between the vectors respectively on the left and on the right of the tensor product in the basis elements of the Liouville space \cite{Cattaneo_2020}, namely  $\ket{i,n}\otimes\ket{j,m}$, where $i,j\in\big\{e,g\big\}$. To find the block-diagonal structure of the Liouvillian, we write it as a matrix in the basis given by the above vectors. Mathematically we can express the complete Liouvillian matrix as a direct sum of the individual blocks as
\begin{equation}
  \label{eq:Liouvillian-block-form}
  \mathcal{L} = \bigoplus_{d}\mathcal{L}_d\,.
\end{equation}
It is noteworthy that the blocks with negative $d$ can be trivially obtained from the blocks with positive $d$ due to the relation $\mathcal{L}_d = \mathcal{L}^*_{-d}$ \cite{Cattaneo_2020}.

This block diagonal form is useful because it allows us to study specific physical phenomena by considering only the blocks where the information about the observables associated with these phenomena is stored. In the case of decoherence, it turns out that we can focus on the $\mathcal{L}_{\pm 1}$ block since only those are needed for the computation of the expectation values $\langle\sigma_x\rangle$ and $\langle\sigma_y\rangle$.  But in order to be able to compute the expectation value, we need to be able to consider the representation for an arbitrary quantum state in the Liouville space, where the Liouvillian is block diagonal. To represent the density matrix in terms of the eigenvectors of the block-diagonal Liouvillian, we follow references \cite{Bellomo_2017} and \cite{Cattaneo_2021_bath}.

Our Liouvillian matrix is in a block diagonal form as depicted in Fig.~\ref{fig:block-form-of-L}. Let us consider the eigenvectors and eigenvalues of some block $\liouv_d$. This block has in total $\sqrt{\text{dim}(\liouv_d)}$ eigenvalues and -vectors. Let us denote the $i$:th eigenvalue of the block $\liouv_d$ by $\lambda_i^{(d)}$ and the corresponding (right) eigenvector by $\kket{v_i^{(d)}}$. Then the (vectorized) density matrix can now be written as a linear combination of these eigenvectors:
\begin{equation}
  \label{eq:rho-liouvillian-eigvec-decomp}
  \kketbig{\rho_\text{Sys}} = \sum_d\sum_i c_i^{(d)}\kketbig{v_i^{(d)}}\,,
\end{equation}
where $d$ goes over all possible blocks of the Liouvillian and $i$ runs from $1$ to $\sqrt{\text{dim}(\liouv_d)}$, denoting the index of the eigenvector of a specific block. The coefficient $c_i^{(d)}$ is the projection of $\kket{\rho_\text{Sys}}$ onto the eigenvector $\kketbig{v_i^{(d)}}$:
\begin{equation}
 \label{eq:coeff-c_i^d}
 c_i^{(d)} = \big\langle\big\langle \tilde{v}_i^{(d)}\kketbig{\rho_\text{Sys}} = \Tr\big[\rho_\text{Sys}\big(\tilde{v}_i^{(d)}\big)^\dagger\big]\,.
\end{equation}
Here $\bbra{\tilde{v}_i^{(d)}}$ denotes the $i$:th \textit{left} eigenvector of the block $\liouv_d$. The eigenvectors are taken to be normalized such that $\bbraket{\tilde{v}_i^{(d)}}{v_j^{(d)}} = \delta_{ij}$.

We know that the quantum state evolves as $\rho(t) = \e^{\liouv t}\rho(0)$. If we now represent the initial state $\rho(0)$ as a linear combination of the eigenvectors of the Liouvillian, as in Eq.~\eqref{eq:rho-liouvillian-eigvec-decomp}, we obtain
\begin{align}
    \kketbig{\rho_\text{Sys}(t)} &= \e^{\liouv t}\kketbig{\rho_\text{Sys}(0)} = \e^{\liouv t}\sum_d\sum_i c_i^{(d)}\kketbig{v_i^{(d)}} \nonumber \\
    &= \sum_d\sum_i c_i^{(d)}\e^{\liouv t}\kketbig{v_i^{(d)}} \nonumber \\
    &= \sum_d\sum_i c_i^{(d)}\e^{\lambda_i^{(d)} t}\kketbig{v_i^{(d)}}\,,
    \label{eq:evolved-rho}
\end{align}
where we used the fact $\liouv\kket{v_i^{(d)}} = \lambda_i^{(d)}\kket{v_i^{(d)}}$. The coefficients $c_i^{(d)}$ give us now the initial conditions. Note that the individual eigenvectors of the Liouvillian are not proper density matrices but their linear combination is, as shown in Eq.~\eqref{eq:rho-liouvillian-eigvec-decomp}.

Now that we have solved for the time evolution of the quantum state in terms of the eigenstates of the Liouvillian, we can consider the expectation value of $\sigma_x$, which is defined as $\langle\sigma_x(t)\rangle = \Tr[\sigma_x\rho(t)] = \bbraket{\sigma_x}{\rho(t)}$. We know what is $\kket{\rho(t)}$ due to Eq.~\eqref{eq:evolved-rho}. The matrix form of $\sigma_x$ is given as
\begin{align*}
  \sigma_x &= \big(\ket{g}\bra{e} + \ket{e}\bra{g}\big)\otimes\mathbb{1}_n \\
  &= \big(\ket{e}\bra{g} + \ket{g}\bra{e}\big)\otimes\big(\ket{0}\bra{0} + \ket{1}\bra{1} + ... + \ket{n}\bra{n}\big)\nonumber \\
           &= \ket{g0}\bra{e0} + \ket{g1}\bra{e1} + \ket{g2}\bra{e2} + ... \\  
           &\quad+ \ket{e0}\bra{g0} + \ket{e1}\bra{g1} + \ket{e2}\bra{g2} + ... \nonumber\,.
\end{align*}
From this we get the vectorized form as
\begin{equation}
  \begin{aligned}
    \Rightarrow \kket{\sigma_x} = &\underbrace{\kket{g0}\otimes\kket{e0} + \kket{g1}\otimes\kket{e1} + ...}_{d=-1} \\
    + &\underbrace{\kket{e0}\otimes\kket{g0} + \kket{e1}\otimes\kket{g1} + ...}_{d=1}\,. \label{eq:vectorized-sigma-x}
  \end{aligned}
\end{equation}
We can notice that the vectorized form contains only the basis elements from the $\liouv_{\pm 1}$ blocks. Therefore, in studying the decoherence of the qubit we can focus only on these two blocks and forget about the rest. Moreover, we do not need to consider all of the values in the $\liouv_{\pm 1}$ blocks as, for example, the basis elements of the form $\kket{g, n+1}\otimes\kket{g, n}$ do not contribute to the decoherence effect. Therefore the number of actual matrix elements that we need to consider is $s^2S^2$, where $s=2$ is the dimension of the qubit Hilbert space and $S$ is the resonator's Hilbert space truncation. This is a significant reduction in the number of values if we compare it to the size of the full Liouvillian matrix, which has $s^4S^4$ elements.

Let's take now the inner product of $\sigma_x$ with the quantum state in order to compute the expectation value:
\begin{align}
  \label{eq:expectation-of-sigmax}
    \langle\sigma_x(t)\rangle &= \bbraket{\sigma_x}{\rho(t)} \nonumber\\
    &= \sum_d\sum_i c_i^{(d)}\e^{\lambda_i^{(d)} t}\bbraketbig{\sigma_x}{v_i^{(d)}} \nonumber \\
    &
    \begin{aligned}
    =\sum_i \Big[&c_i^{(1)}\e^{\lambda_i^{(1)} t}\bbraketbig{\sigma_x}{v_i^{(1)}} \\
    &+ c_i^{(-1)}\e^{\lambda_i^{(-1)} t}\bbraketbig{\sigma_x}{v_i^{(-1)}}\Big]\,.
    \end{aligned}
\end{align}
From the equivalence between the Liouvillian blocks, $\liouv_d = \liouv^*_{-d}$, it follows that $c_i^{(d)} = c_i^{(-d)*}$ and $\bbraket{\sigma_x}{v_i^{(d)}} = \bbraket{\sigma_x}{v_i^{(-d)}}^*$. Also the eigenvalues $\lambda_i^{d}$ follow the same kind of logic, and thus they can be written as $\lambda_i^{(\pm d)} = \text{Re}[\lambda_i^{(d)}] \pm \i\text{Im}[\lambda_i^{(d)}]$. Using these results in Eq.~\eqref{eq:expectation-of-sigmax} allows us to write it in the following way:
\begin{equation}
  \label{eq:expectation-of-sigmax2}
  \begin{aligned}
  \langle\sigma_x(t)\rangle = \sum_i \Big[&c_i^{(1)}\bbraketbig{\sigma_x}{v_i^{(1)}}\e^{\text{Re}[\lambda_i^{(1)}] t}\e^{\i\text{Im}[\lambda_i^{(1)}]t} \\
  &+ c_i^{(1)*}\bbraketbig{\sigma_x}{v_i^{(1)}}^*\e^{\text{Re}[\lambda_i^{(1)}] t}\e^{-\i\text{Im}[\lambda_i^{(1)}]t}\Big]\,.
  \end{aligned}
\end{equation}

The factor $c_i^{(1)}\bbraket{\sigma_x}{v_i^{(1)}}$ is a complex number that keeps track of the initial conditions. Denoting $\bbraket{\sigma_x}{v_i^{(1)}} \equiv p_i$ in Eq.~\eqref{eq:expectation-of-sigmax2} leads us to the final form for the expectation value of $\sigma_x$ in Eq.~\eqref{eq:expectation-of-sigmax3}. This equation decomposes the time evolution of $\sigma_x$ into a sum of different modes. The oscillation of the different modes is driven by the imaginary part of the corresponding eigenvalue, while the real part is responsible for the decay of the mode. Thus we can see that if some eigenvalue is purely real, that induces pure decay for the corresponding mode without any oscillations. An opposite case is a purely imaginary eigenvalue, where we see no decay and only oscillations of the coherence. It can be shown that any eigenvector with zero eigenvalue is a steady state of the dynamics \cite{Albert_2014,minganti2018spectral}. However, the blocks $\mathcal{L}_{\pm 1}$ have no zero eigenvalues, while the latter are usually found in the block $\mathcal{L}_0$ only \cite{Cattaneo_2020}.

%% file: appendix6.tex
\section{Effect of the initial state of the resonator to coherence decay} \label{appendix6}

In the following figures we explain the effect of the initial state of the resonator on the qubit decoherence. The parameters in these figures have been chosen so that the timescale of the dynamics is sufficiently fast that we are able to observe and distinguish the individual oscillations of the qubit coherences (solid blue (light gray in print) line).

\begin{figure}
    \centering
    \includegraphics[width=\linewidth]{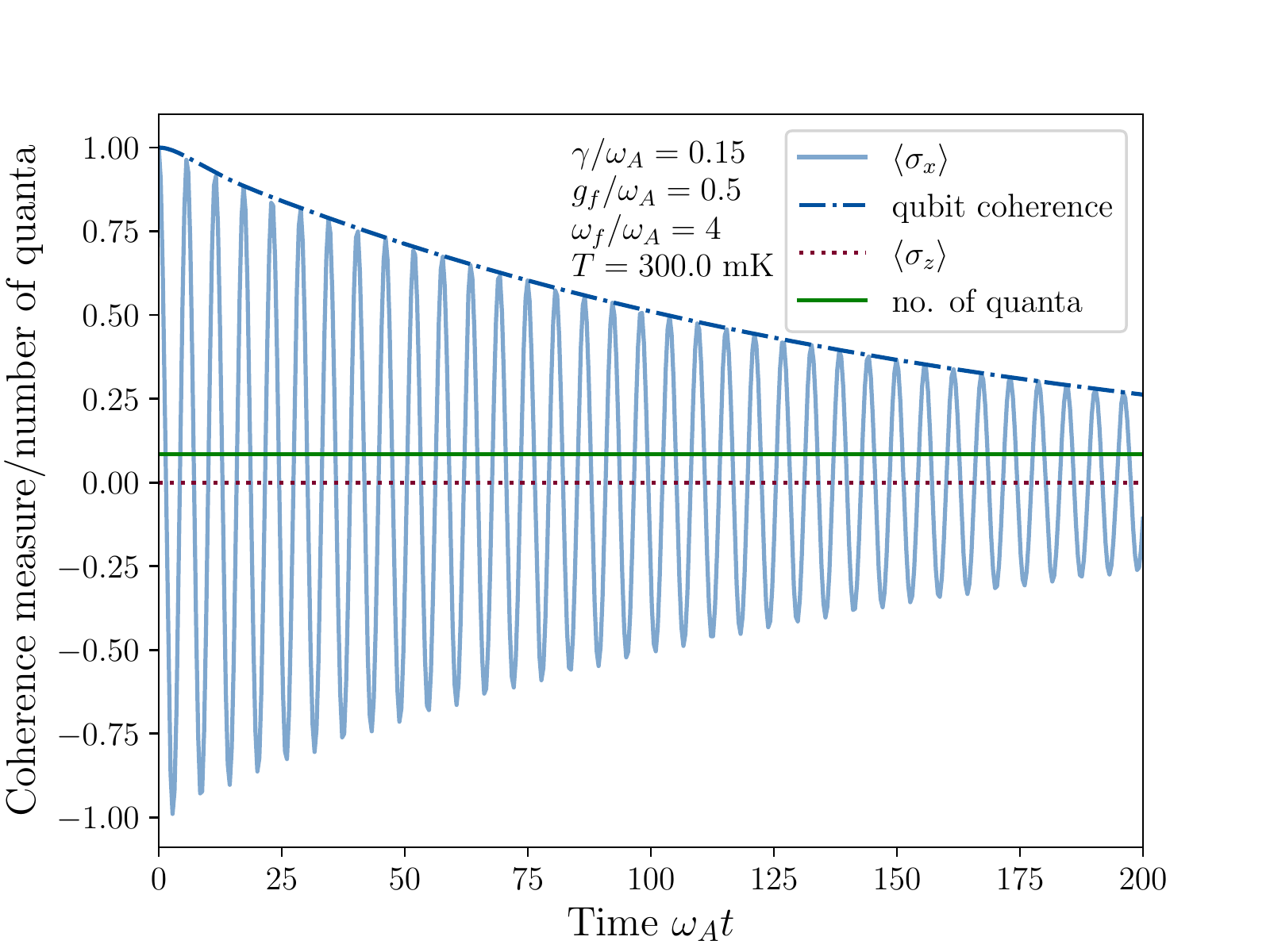}
    \caption{System evolution when the resonator is initialized in the thermal state. In this scenario, the qubit experiences decoherence at a (for all practical purposes) constant rate. The almost exponential decay is well-captured by the decay rate in Eq.~\eqref{eq:T2-timescale}.}
    \label{fig:thermal}
\end{figure}

\begin{figure}
    \centering
    \includegraphics[width=\linewidth]{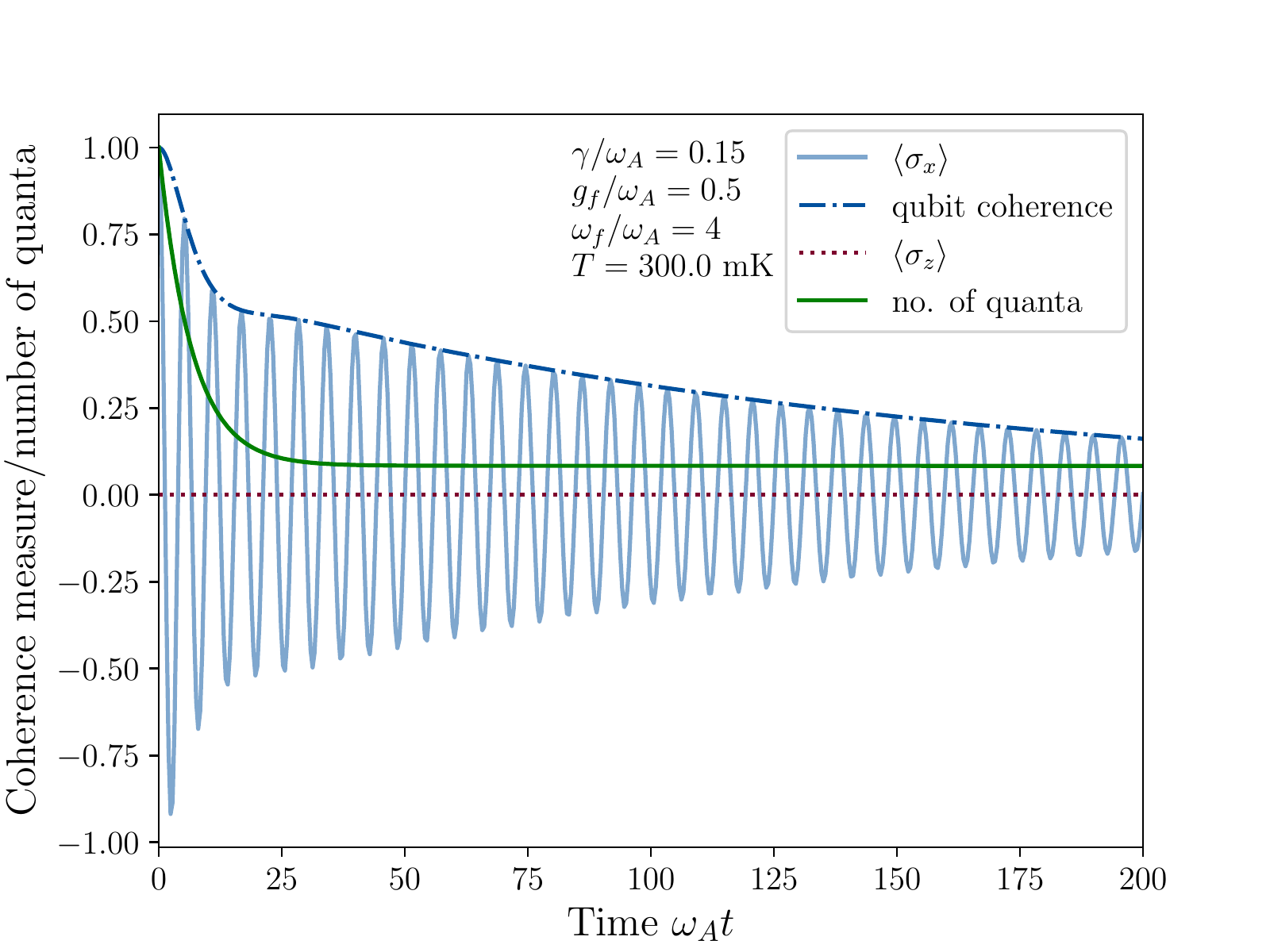}
    \caption{System evolution when the resonator is initially in a coherent state and the resistor temperature is $T=300$mK. While the resonator decays towards the steady state, the qubit experiences faster decoherence due to a faster eigenmode of the Liouvillian being present. After the reduced state of the resonator has reached the thermal state, the qubit continues decohering at a slower rate determined by Eq.~\eqref{eq:T2-timescale}.}
    \label{fig:coherent}
\end{figure}

\begin{figure}
    \centering
    \includegraphics[width=\linewidth]{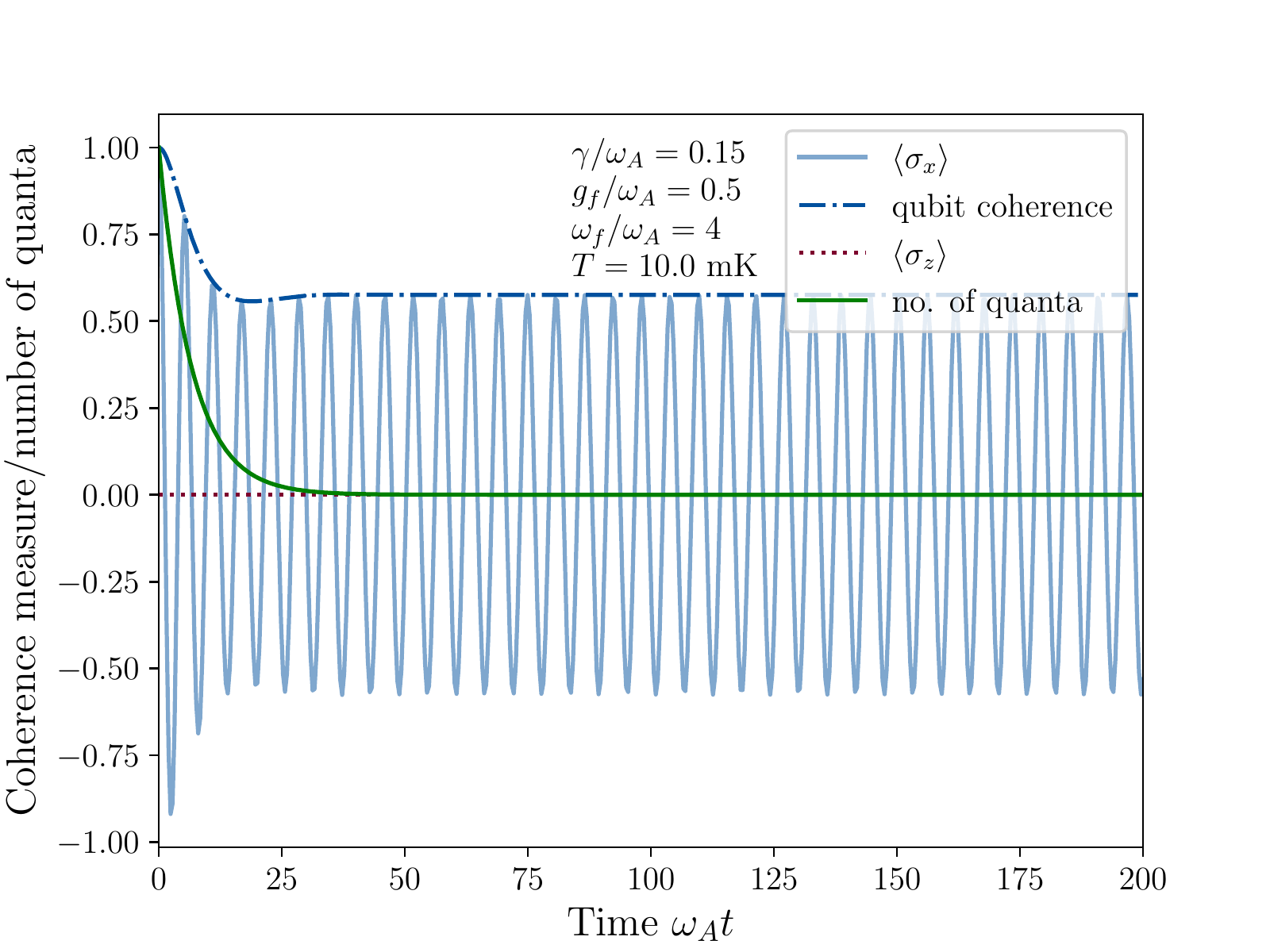}
    \caption{System evolution when the resonator is initialized in a coherent state and the resistor temperature is $T=10$mK. During the decay of resonator state towards the thermal state the qubit experiences rapid decoherence. Once the resonator has reached its steady state the rate of qubit decoherence becomes vanishingly small due to very cold bath temperature.}
    \label{fig:coherent cold}
\end{figure}

As was argued in Sec.~\ref{decoher for disp transmon} and appendix \ref{appendix5}, the time evolution of the quantum state of the qubit can be presented in terms of the evolution of the different eigenmodes of the Liouvillian superoperator (see Eqs.~\eqref{eq:expectation-of-sigmax3} and \eqref{eq:evolved-rho}). Some eigenmodes have faster decay than others and, in general, the faster modes are present when the resonator is not in its steady state, but on its way there. This can be seen by comparing Figs.~\ref{fig:thermal} and \ref{fig:coherent}. In Fig.~\ref{fig:thermal} the qubit decoheres (coherence measure plotted as the blue dash-dot line) in an almost exponential fashion when the resonator starts out in its steady state, the thermal state. The fact that the resonator is in a thermal state can be seen by looking at the solid green line in Fig.~\ref{fig:thermal}, which indicates the number of quanta in the resonator and stays at a constant value, which is given by the Bose-Einstein distribution for that temperature. The decay rate for the coherence measure of the qubit is well described by Eq.~\eqref{eq:T2-timescale}.

In Fig.~\ref{fig:coherent}, the resonator is initialized in a coherent state and thus it too experiences decay, which is seen as the change in the number of quanta in the resonator. We can see that while the resonator is decaying towards the thermal state, the qubit experiences a faster coherence decay, up until to the point where the resonator has finally reached its steady state. After that, the qubit continues to decohere at an exponential fashion with the decay rate given once again by Eq.~\eqref{eq:T2-timescale}. 

In Fig.~\ref{fig:coherent cold} the resonator starts out in the coherent state as in Fig.~\ref{fig:coherent} but the temperature is lowered to $T=10\,\mathrm{mK}$. In this case the qubit decoheres at a quick rate in the beginning as the resonator is moving towards its steady state. But in this case the temperature is so low that when the resonator reaches its thermal state, the qubit decoheres so slowly that it is not visible in the timescale in Fig.~\ref{fig:coherent cold}. It should be noted that at $T=0\,\mathrm{K}$ the qubit will stop decohering altogether once the resonator reaches its thermal state $\ket{0}\bra{0}$, as explained at the end of Sec.~\ref{decoher for disp transmon}.